\documentclass[letterpaper,useAMS,usenatbib]{mn2e}
\usepackage{comment}
\usepackage{pdflscape}

\usepackage{amssymb}   
\usepackage{amsmath}   
\usepackage{amssymb}   

\usepackage{bm}
\newcommand{\bs}{\boldsymbol}

\usepackage{xspace}
\usepackage{graphicx}
\usepackage{verbatim}

\usepackage[T1]{fontenc}
\usepackage{aecompl}

\newcommand{\be}{\begin{equation}}
\newcommand{\ee}{\end{equation}}

\title[Narrow line lensing]{Detection of substructure with adaptive optics integral field spectroscopy of the gravitational lens B1422+231}

\author[Nierenberg et al.]{
A.~M.~Nierenberg$^{1,2}$, 
T.~Treu$^{1,3}$, 
S.~A.~Wright$^{4}$,
C.~D.~Fassnacht$^{5}$,
M.~W.~Auger$^{6}$
\medskip\\
$^1$Department of Physics, University of California, Santa Barbara, CA 93106, USA\\
$^2${\tt amn01@physics.ucsb.edu}\\
$^3$ Packard Fellow \\ 
$^4$ Dunlap Institute for Astronomy and Astrophysics, 50 St. George St., Toronto, ON, M5S 3H4, Canada \\ 
$^5$ Department of Physics, University of California, Davis, CA 95616, USA \\ 
$^6$ Institute of Astronomy,University of Cambridge, Madingley Road, Cambridge, CB30HA, UK\\ 
}

\begin{document}
\date{Accepted for publication in MNRAS}
\pagerange{\pageref{firstpage}--\pageref{lastpage}}\pubyear{2013}

\maketitle           

\label{firstpage}

                      
\begin{abstract}
Strong gravitational lenses can be used to detect low mass subhalos, based on deviations in image fluxes and positions from what can be achieved with a smooth mass distribution. So far, this method has been limited by the small number of (radio-loud, microlensing free) systems which can be analysed for the presence of substructure. Using the gravitational lens B1422+231, we demonstrate that adaptive optics integral field spectroscopy can also be used to detect dark substructures.  We analyse data obtained with OSIRIS on the Keck I Telescope, using a Bayesian method that accounts for uncertainties relating to the point spread function and image positions in the separate exposures. The narrow-line [OIII] fluxes measured for the lensed images are consistent with those measured in the radio, and show a significant deviation from what would be expected in a smooth mass distribution, 
 consistent with the presence of a perturbing low mass halo. Detailed lens modelling shows that image fluxes and positions
 are fit significantly better when the lens is modelled as a system containing a single perturbing subhalo in addition to the main halo, rather than by the main halo on its own, indicating the significant detection of substructure.
The inferred mass of the subhalo depends on the subhalo mass density profile: the 68 \% confidence interval for the perturber mass within 600 pc are: 8.2$^{+0.6}_{-0.8}$ , 8.2$^{+0.6}_{-1}$ and 7.6$\pm0.3$ $\log_{10}$[M$_{\rm{sub}}$/M$_\odot$] respectively for a singular isothermal sphere, a pseudo-Jaffe, and an NFW mass profile. This method can extend the study of flux ratio anomalies to virtually all quadruply imaged quasars, and therefore offers great potential to improve the determination of the subhalo mass function in the near future.
\end{abstract}

\setcounter{footnote}{1}

\section{Introduction}
\label{sec:intro}

A fundamental prediction of the $\Lambda$ Cold Dark Matter ($\Lambda$CDM) model is that a Milky Way mass halo should be surrounded by thousands of subhalos. However, only about twenty are observed \citep[e.g.][]{Klypin++99, Moore++1999, Strigari++07, Toll++08, Koposov++08}. Significant work in the past decade has gone into understanding this failure in a theory that has otherwise been successful at reproducing the observed power spectrum of structure in the Universe over an enormous range of distance, mass and time scales \citep[e.g.][]{Planck++13,Komatsu++11}.

If CDM is correct, there must be a large number of subhalos which do not retain enough gas \citep{Papastergis++11} or form enough stars to be detected. A variety of baryonic processes are suspected to lead to dark subhalos. In fact, star formation in satellite galaxies has been suggested to be quenched by a variety of internal and external process such as UV heating during reionization, tidal and ram-pressure stripping by the central galaxy, supernova feedback, and stellar winds \citep[e.g.][]{Lu++12, Menci++12, Guo++11,Bullock++2000,Benson++2002, Somerville++02, Kravtsov++04, Kaufmann++08, Maccio++10,Springel++10,Zolotov++12}. Significant uncertainty as to the relative importance of various star formation processes remains, with numerous models reproducing the luminosity function of Milky Way satellites while failing to match the satellite luminosity function in different host mass and redshift regimes \citep{Nierenberg++13}. An alternative explanation for the lack of satellites is that dark matter subhalos are not as numerous as predicted by $\Lambda$CDM, as in the case of warm dark matter models \citep{Nierenberg++13}.

There are also discrepancies between the subhalo mass profiles predicted by dark matter only CDM simulations and what is observed in Local Group dwarf galaxies. \citet{Boylan-Kolchin++11, Boylan-Kolchin++12}, pointed out that the most luminous Milky Way satellites reside in halos which have much lower circular velocities than what is predicted by simulations. Baryonic processes could account for a flattening of the central density profile in the more massive satellite galaxies \citep{Zolotov++12}. Alternatively, self-interacting cold dark matter (SIDM) has been proposed to explain the flattening \citep[e.g.][]{Rocha++13}.


The only way to fundamentally distinguish between these various dark matter models is with a direct measurement of the subhalo mass function which does not rely on an understanding of the baryonic physics in the subhalos. In the Local Group, new work has shown the possibility that dark subhalos may be detected via their interactions with tidal streams \citep{Carlberg++12}, or the HI disk of the Milky Way \citep{Chakrabarti++09}. However, these methods are currently limited to low redshifts, and rely on detailed modelling of baryonic structures. 

Outside of the low redshift Universe, gravitational lensing can be used to measure the subhalo mass function at a range of redshifts, without requiring detailed modelling of baryonic physics. Weak gravitational lensing can be used to measure the masses of massive ($\sim 10^{11.5}$M$\odot$) satellites of galaxy groups \citep{Li++14}, while strong gravitational lensing is sensitive to lower mass subhalos. In strong gravitational lensing, a background source is multiply imaged by a foreground deflector, with the image time-delays, positions, and magnifications depending sensitively on the gravitational potential of the deflector, and its first and second derivatives respectively. Precise astrometry and photometry of lensed images have been shown to be powerful ways to measure the mass of low-mass subhalos in proximity to lensed images \citep[e.g.][]{Mao++98,M+M01,Dalal++02,Koo05, Ama++06,McKean++07, K+M09, MacLeod++09,Treu++10,Vegetti++10a,Vegetti++12,Fadely++12}.

There are two main requirements for a strong gravitational lens to be suitable for the detection of substructure. First, the background source must be sufficiently large that it is not significantly magnified by stars in the plane of the lens galaxy. For typical lens configurations, stars have Einstein radii of order microarcseconds, thus the background source must be significantly larger than microarcseconds in apparent size. Secondly, the lensed images must contain enough information to constrain the lensing `macromodel', i.e., the smooth mass distribution of the main deflector. 

Traditionally, these restrictions have limited the study of substructure to the seven known radio-loud quasar sources which are quadruply imaged (quad).  In a seminal work, \citet{Dalal++02} demonstrated that these systems contained a fraction of mass in substructure which was broadly consistent with predictions from CDM simulations, albeit with large uncertainties. Progress with this technique has been limited by the small number of known suitable systems.

An alternative method of using gravitational lensing to study substructure is gravitational imaging, proposed by \citet{Koo05}, in which subhalos are detected via astrometric perturbations of lensed galaxies. This method has been successfully applied to HST and AO imaging \citet{Vegetti++12, Vegetti++10b}, and simulations indicate it will also work for ALMA data \citep{Hezaveh++13}. Recently \citet{Vegetti++14} analysed a sample of 11 lenses, finding a fraction of substructure consistent with numerical simulations and results from analyses of four image quasar lenses.
The mass sensitivity of this method is determined by the signal to noise ratio and resolution of imaging, as well as the intrinsic morphology of the lensed galaxy. 
Current measurements with Keck AO are sensitive to substructure with masses of $\sim 10^{7.5}$M$_\odot$ and larger, while measurements with the Next Generation Adaptive Optics on Keck, the Thirty Meter Telescope, Very Long Baseline Interferomety may allow the sensitivity of this method to reach masses of $\sim 10^6$M$_\odot$.

 Strongly lensed \emph{narrow-line} quasar emission provides an alternative means of measuring the subhalo mass function \citep{Moustakas++03}.
 Quasar narrow-line emission at low redshift is observed to be typically extended over tens to hundreds of parsecs depending on the source luminosity \citep{Bennert++02}, which corresponds to milliarcseconds for typical source redshifts, well above the microlensing scale. A benefit to studying narrow-line emission is that many more quasars have detectable narrow-line emission as opposed to radio emission. This makes this method ideal for measuring the subhalo mass function in the thousands of quad quasar lenses which are expected to be discovered based on their optical properties in ongoing and future surveys including PANSTARRS, DES and LSST \citep{Oguri++10}.

While promising, spatially resolved spectroscopy of narrow line emission of quasar lenses has proven difficult to attain. \citet{Sluse++12a} and \citet{Guerras++13} have measured lensed image spectra for pairs of quasar images, and for wide image separation double lenses, and observed differential lensing between the broad and continuum emission, which is evidence for microlensing. In order to study the lensing signal of substructure as discussed before, it is necessary to measure the spectra of images individually. \citet{Metcalf++04} and \citet{Keeton++06}, achieved sufficient spatial resolution in order to measure the broad emission in individual quad quasar images, but neither of these studies detected narrow emission with sufficient signal to noise ratio. 

In a pioneering work, \citet{Sugai++07} used integral field unit (IFU) optical spectroscopy in order to measure narrow [OIII] fluxes in the gravitational lens RXJ1131. The relatively wide separation between the lensed images ($\sim 1\farcs$0) made it possible to spatially resolve the lensed images, and place an upper limit on the mass of perturbing substructure. Interestingly, they found that unlike the continuum and broad fluxes, the [OIII] image fluxes did not deviate significantly from a smooth model prediction without substructure. In order to apply this method to a larger sample of systems and to probe lenses with configurations with smaller image separations (such as fold lenses) it is essential to obtain higher spatial resolution via HST \citep{Keeton++06}, or adaptive optics from the ground.
 In this work, we demonstrate for the first time that subhalos can be detected using strongly lensed narrow-line quasar emission of a high redshift (z=3.6) quasar by combining adaptive optics with integral field spectroscopy in order to obtain spatially resolved spectra of individual lensed images.

In this paper we present a measurement of narrow-line lensing with
unprecedented accuracy using the OH Suppressing Infra-Red Imaging
Spectrograph \citep[OSIRIS][]{Larkin++06} with laser guide star
adaptive optics (LGS-AO) \citep{Wizinowich++06} at Keck.  We choose as
initial case to study the famous system B1422+231
\citep{Patnaik++92}. Previous extensive studies across many bands of
photometry (including radio) enable us to compare our measurement with
more traditional methods. 

The structure of the paper is as follows: In \S \ref{sec:data} we present our observations. In \S  \ref{sec:imExtract} we describe our Bayesian method for combining multiple exposures to get sub-pixel sampling of the images and extract spectra optimally. In \S \ref{sec:spectra} we present the extracted spectra and show fits to the broad and narrow line components in each of the lensed images, and the inferred fluxes from each. In \S \ref{sec:lensmodels} we discuss the gravitational lens models we use to fit the observed image fluxes and positions. In \S \ref{sec:results} we present results from a gravitational lens model assuming a single perturbing subhalo. In \S \ref{sec:discussion} we discuss future prospects for this method. Finally in \S \ref{sec:summary}, we summarise our results. Throughout this paper, we assume a flat $\Lambda$CDM cosmology with $h=0.7$ and $\Omega_{\rm m}=0.3$.  All magnitudes are given in the AB system \citep{Oke++1974} unless otherwise stated.

\section{Observations}
\label{sec:data}

We observed the gravitationally lensed quasar B1422+231 on June 10 2012 at the W. M. Keck 10 m telescopes, using OSIRIS with LGS-AO corrections. 
To maximise the signal-to noise ratio we used the largest available pixel scale which is $0\farcs1$. We used broad-band K ($\lambda=$2.17 microns) filter which provided a spatial field of view of $\sim 1\farcs6$x$6\farcs4$ and a spectral range of 416 nm in the observed frame. At the time of our observations, the instrument had recently been moved and the full calibration had not yet been performed. As a result we were limited to a bigger wavelength range and smaller field of view that made it possible to only fit three of the four images in the field of view.  We observed the system using 300 s exposures, dithering along the long axis of the field of view, with sub-pixel offsets in order to recover more spatial information, for a total of 2700s of integration. We used the OSIRIS data reduction pipeline to turn the raw CCD images from each exposure into rectified, telluric corrected, and sky emission-subtracted data `cubes' in which the spatial information of the images appeared in the x-y plane of the images, and the wavelength varied along the third dimension.

\section{Data Reduction}
\label{sec:imExtract}

Our goal is to measure integrated line fluxes separately for each lensed image, taking into account all uncertainties regarding image positions, image deblending, and PSF properties. We do this in a three step process.

\begin{figure}
\centering
\includegraphics[scale=0.65,trim = 200 300 180 10, clip = true]{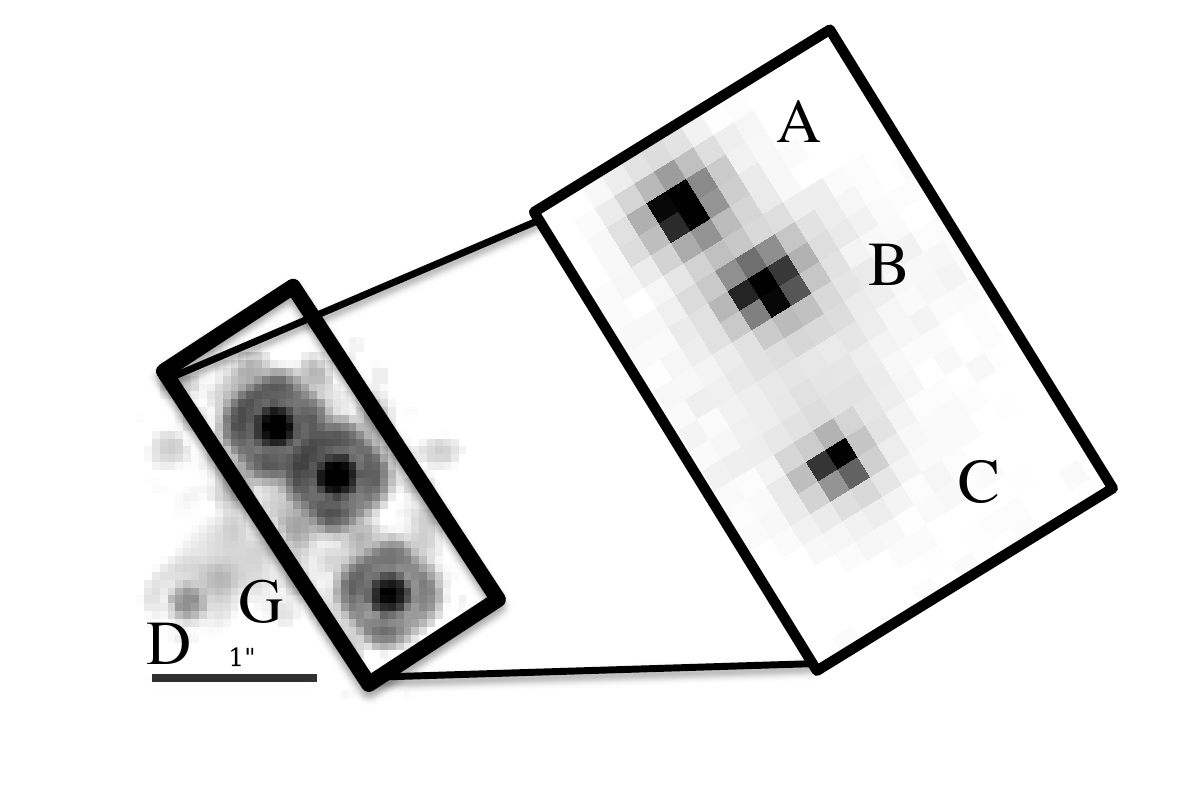}\\
\includegraphics[scale=0.55,trim = 0 0 0 0, clip = true]{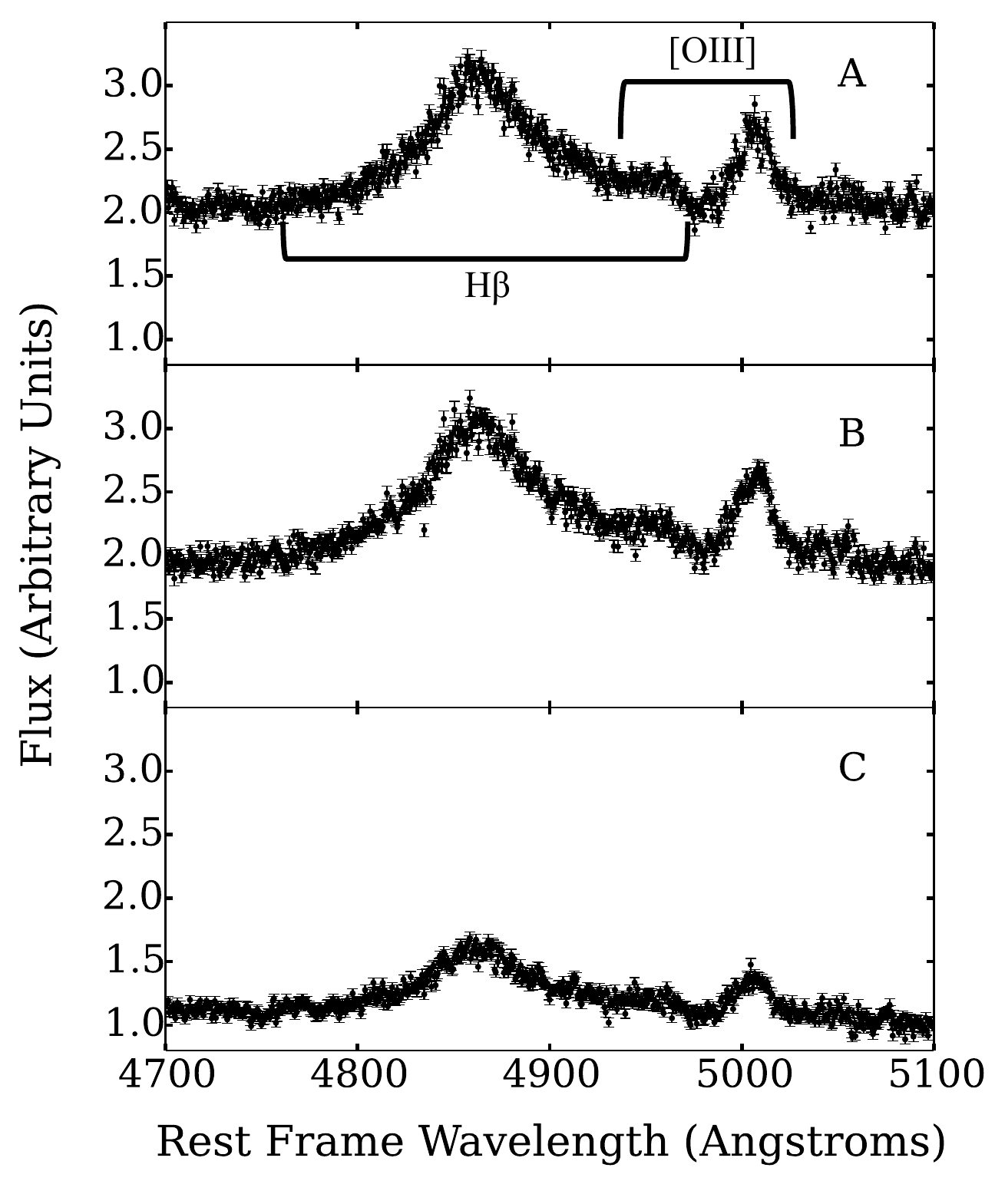}
\caption{Upper Left: HST NICMOS F160W image. Upper Right: Single 300s K band exposure of target using OSIRIS with adaptive optics at Keck, we were restricted to the small field of view due to instrument difficulties. Lower panels: The final extracted spectra for each of the images.}
\label{fig:spectra}
\end{figure}

First, we infer the properties of the PSF and image positions in `white' images, which we created by taking the variance-weighted mean of the data cube over the wavelength region dominated by the broad and narrow line emission between rest frame 4700 - 5100 \AA. These white images are effectively narrow band images of the system. Figure \ref{fig:spectra} shows the white image for a single exposure. 

Next, we use a Bayesian inference in order to model the system as three point sources (e.g., the three observed lensed-QSO images).   The centre of lens galaxy is not within the field of view (Figure \ref{fig:spectra}), and is $\sim3$ magnitudes fainter than the QSO images, thus we do not include its contribution to light in the image.
The model contains a set of global parameters that are the same for all exposures (the offsets between the three sources and their relative fluxes) and a set of parameters that vary from exposure to exposure (the sky level, the absolute location of the images, the total flux of the three images, and properties describing the shape of the PSF for each exposure). The PSF is modelled as two concentric Gaussians to represent the seeing halo and the AO corrected diffraction core. The inner diffraction core is fixed to be circular and to have a full width at half maximum (FWHM) of 0\farcs065, based on the size of the telescope and the wavelength of observation \citep[see also][]{vanDam++06}. The seeing halo Gaussian FWHM, ellipticity, position angle, and amplitude relative to the diffraction core Gaussian are free parameters. We run an Markov Chain Monte Carlo (MCMC) sampler for at least ten thousand steps in order to explore the full posterior probability distribution function, including degeneracies between the parameters.  We tested this method by simulating images with the same image separations and signal-to-noise ratio as our data set, and found that we could recover the image positions to an accuracy of better than 5 mas, and the image fluxes to percent level accuracy. 

The resulting measured image positions are given in Table
\ref{tab:imposflux}.  The relative image positions positions
are consistent with radio and HST measurements of the system.
Our observations occurred during excellent
conditions, with typical Strehl $\sim$ 0.2-0.3, and seeing of $\sim
0\farcs3$

Finally, we extract the spectra using the inferred PSF properties and image positions, relying on the fact that the PSF does not vary significantly over the short wavelength range we are considering ($\sim 180$ nm observed).  We draw sets of parameters describing the PSF and image positions from the MCMC chains obtained in the previous step. Then, for each set of parameters, for each wavelength slice in the data cube, we perform a $\chi$-squared optimisation via linear inversion in order to find the best fit amplitudes for the lensed images simultaneously for all exposures, allowing for variations in the background flux as a function of wavelength and exposure. The final extracted spectrum for each image is then given by the mean and standard deviation of the best fit image fluxes in each wavelength slice after iterating over all of the parameter draws. This method allows us to quickly and robustly incorporate uncertainties in our inference of the image positions and PSF parameters directly into our spectral extraction. Image spectra are plotted in Figure \ref{fig:spectra}.

\begin{table*}
\scriptsize\begin{tabular}{llllll}
\hline \hline
Component & dRa & dDec & [OIII] flux & broad H-$\beta$  & continuum \\
\hline 

A & 0.387 $\pm 0.005$  & 0.315   $\pm 0.005$& 0.88  $\pm$ 0.01         & 4.67 $\pm$ 0.06   & 46.3  $\pm$ 0.1  \\ 
B & 0     $\pm 0.005$  & 0   $\pm 0.005$        & 1.00 $\pm$0.01          & 5.2 $\pm 0.1$      &  44.6  $\pm 0.1 $   \\ 
C & -0.362 $\pm 0.005$ &-0.728 $\pm 0.005$ & 0.474  $\pm$ 0.006   & 2.5 $\pm 0.1$      & 24.1 $\pm 1$   \\ 
D & 0.941  $\pm 0.01$  & -0.797  $\pm 0.01$    & -          & -           & - \\ 
G & 0.734 $\pm 0.01$ &-0.649 $\pm 0.01$ & -               &-              &- \\
\hline \hline
\end{tabular}
\caption{Image positions and fluxes for images A B and C measured with OSIRIS. Fluxes are in units of the [OIII] flux of image B which was measured to be $1.4\pm 0.2 \times 10^{-14}$erg/(s cm$^2$) by \citet{Murayama++1999}. Distances are in units of arcseconds. The position for image D comes from averaging the offsets from radio observations by \citet{Patnaik++1999}  (see Equation 1), while the position for the main lens galaxy (G) comes from averaging the offsets from HST observations from the CASTLES website. \label{tab:imposflux}}
\end{table*}

\section{Integrated Line Fluxes}
\label{sec:spectra}

Having extracted spectra for each of the lensed images, the next step in our analysis is to measure the integrated line fluxes by modelling the spectra as a sum of narrow H-$\beta$ and [OIII], broad H-$\beta$ and continuum emission\footnote{We also tested for the presence of broad iron nuclear emission, but did not find a significant contribution, and thus omitted it from the final analysis.}.

Our data give us three independent measurements of the quasar
spectrum, which we expect to be magnified relative to each other due
to gravitational lensing. We model the broad H-$\beta$ emission and
narrow emission lines as being composed as linear sums of 5th and 3rd
order Gauss Hermite polynomials respectively, while the continuum is
modelled as a power law.  We allow more flexibility in the fit to 
the broad emission line to account for winds which may cause 
significant asymmetries in the line profile.
The model imposes the same narrow line FWHM
and centroid for all 3 lensed QSO images.  The broad and narrow
emission lines are allowed to be redshifted relative to each other to
account for winds.

\begin{figure}
\centering
\includegraphics[scale=0.45]{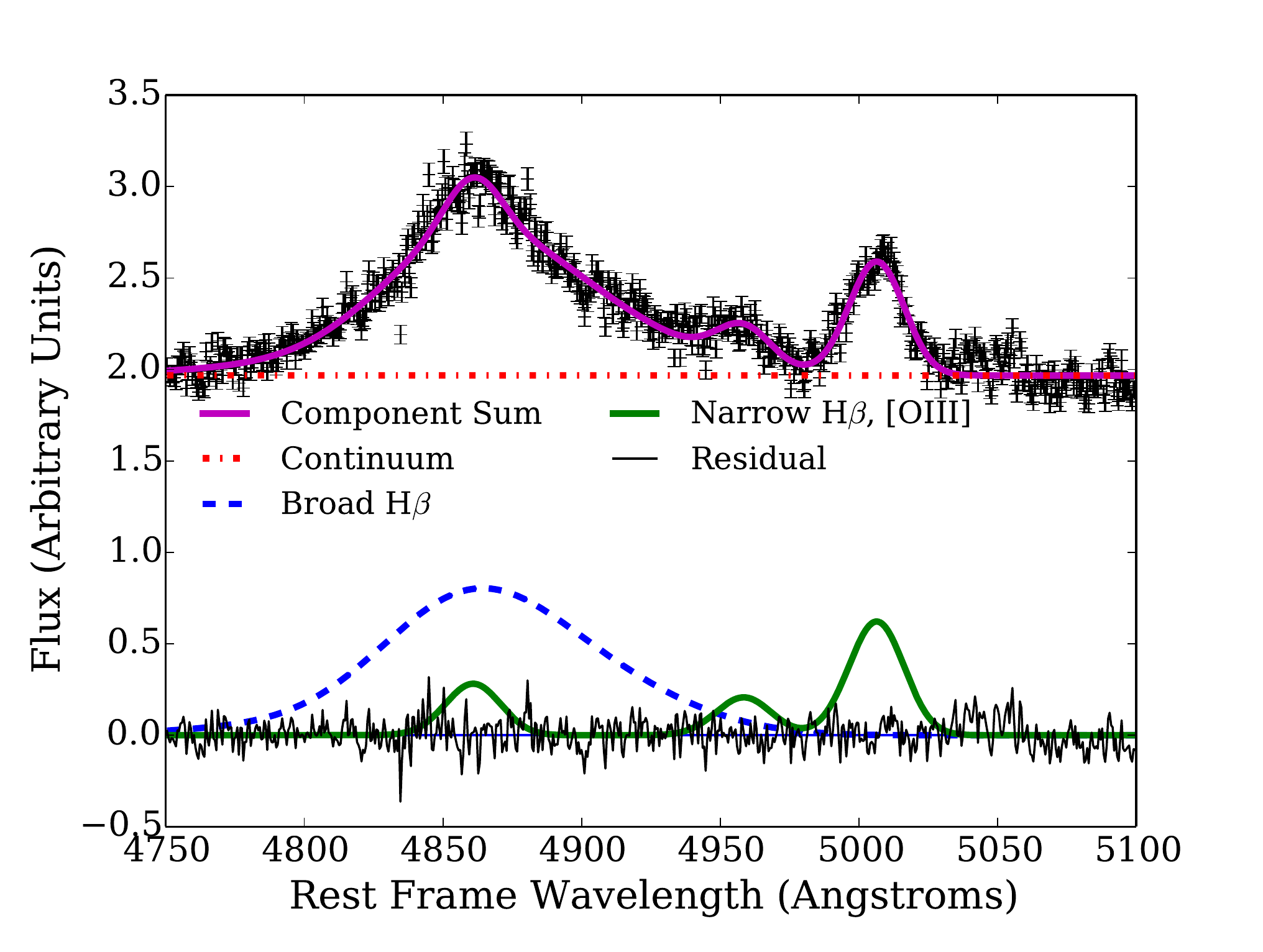}
\caption{Demonstration of best fitting model spectrum to the spectrum of image B, as well as each of the separate line components.\label{fig:specDecompose}}
\end{figure}

The amplitudes of the continuum, narrow, and broad emission lines are all free to vary between the lensed images. The broad line width and shape (i.e. the width and amplitudes of the component Gauss-Hermite polynomials which are summed to model the broad line emission), and the continuum slope are also inferred separately for each spectrum to account for the differential effects of microlensing and intrinsic variability, which 
will more strongly affect the continuum and high-velocity tails of the broad emission line as these are emitted from the smallest physical area. Figure \ref{fig:specDecompose} shows an example of the decomposition of the spectrum into the four components for image B. Uncertainties in the decomposition are incorporated into the final calculated line fluxes by drawing 1000 random samples from the posterior probability distribution of the spectral parameters and re-computing the line fluxes for each sample draw. The integrated line fluxes with measurement uncertainties for each image are given in Table \ref{tab:imposflux}.  In the case of [OIII], the fluxes of both lines are summed together.


\subsection{Astrometry of image D and the main lens galaxy}

Although our measurement with OSIRIS only contains information for image fluxes and positions for three of the four images of the system, we incorporate information on the position of the fourth image from previous radio imaging by \citet{Patnaik++1999}. The position of the fourth image aids in the constraint of the model for the main deflector, and is not expected to vary relative to the other images over time.  We take the position of image D to be the mean position after applying the offsets measured by \citet{Patnaik++1999} between the three measured image positions: 

\be
\bm{x}_D = 1/3\sum_{i=A,B,C}\bm{x}_{i} - (\bm{x}_{i,p} - \bm{x}_{D,p}), 
\ee
where $\bs{x}_{p,i}$ represents position measurements of the $i^{th}$ image by \citet{Patnaik++1999}, and $\bs{x}_i$ is the corresponding measurement in this work. The uncertainty in position D is the standard deviation of the positions. The flux of image D is not used as a constraint in the lens modelling.

Similarly, we use the relative position of the lens galaxy in HST imaging from the CfA-Arizona Space Telescope LEns Survey (CASTLES) website\footnote{\tt www.cfa.harvard.edu/castles/} in order to constrain the centroid of the main deflector in the lens model.

\section{Gravitational Lens Modelling}
\label{sec:lensmodels} 

Having obtained the integrated narrow-line fluxes and image positions, we can now test for the presence of substructure in the system using gravitational lens modelling. As noted in the introduction, the positions and fluxes of the lensed narrow-line emission images are determined by the first and second derivatives respectively of the gravitational potential of the deflector, and therefore can be used to recover the mass distribution of the perturber as well as the main deflector. 

We begin in Subsection \ref{sec:smooth} by modelling the lens as
containing a single main deflector with no substructure in an external
shear field, and find the best fit mass model parameters given our
image fluxes and positions using {\tt gravlens} by
\citet{Keeton++01a,Keeton++01b}. As with previous studies, we find
that this mass model provides a poor fit to the image fluxes \citep{Mao++98, Keeton++01, Bradac++05, Sluse++12b, Dobler++06}.  In
Subsection \ref{sec:perturbed}, we explore the effects of adding a
single perturber to the system with varying mass profiles.

\subsection{Singular Isothermal Ellipsoid Deflector}
\label{sec:smooth}

Following numerous previous studies, we model the deflector as a singular isothermal ellipsoid (SIE) which has been shown to provide a good fit to the macroscopic mass distribution of lens galaxies \citep{Treu++10}. The SIE has a density distribution given by:
\be
\rho(r) = \frac{\rho_o}{r^2} = \frac{\sigma^2_v}{2 \pi r^2 G} = \frac{b}{2 \pi r^2}\left(\frac{c^2}{4\pi G}\frac{D_S}{D_{LS}D_L}\right)
\label{eq:sis}
\ee
Where $\sigma_v$ is a velocity dispersion, $b$ is the Einstein radius of the lens, and $D_S$, $D_{LS}$ and $D_L$ are the angular diameter distances from the observer to the source, from the lens to the source and from the observer to the lens respectively. In {\tt gravlens}, the radius is in elliptical coordinates such that:
\be
r = \sqrt{x^2(1-\epsilon)+y^2(1+\epsilon)},
\ee
where $\epsilon$ is related to the axis ratio $q$ by $q^2 = (1-\epsilon)/(1+\epsilon)$. This means that the Einstein radius is defined along the intermediate axis between the major and minor axes.

As is customary, we also allow for the presence of external shear $\gamma$ in
the direction $\theta_\gamma$ in order to account for the fact that
B1422+231 is a member of a group of
galaxies \citep[e.g.][]{Kundic++97}.

We determine the lens model parameters by first using the {\tt gravlens} optimisation routine in order to find the maximum likelihood parameters for the central deflector given the observed image positions and narrow-line flux ratios. We then use {\tt emcee hammer} \citep{Mackey++13}, which is a Markov Chain Monte Carlo algorithm which is efficient at exploring highly degenerate parameter spaces, in order to determine uncertainties in the parameter values. We initialise the proposal values in a small region in parameter space around the best fit solutions from {\tt gravlens}, and then for each proposed set of host properties, we use {\tt gravlens} in order to optimise the source position and compute the $\chi^2$ for the resulting image positions and magnifications.

The inferred median and one sigma uncertainties for the lens model parameters are given in Table \ref{tab:modelPars}. 
The corresponding model prediction for the image fluxes and positions are given in Table \ref{tab:modposflux}. The model provides an excellent fit to the image positions, matching them within the measurement uncertainties, however the predicted image flux ratios deviate significantly from the observation, so that the typical model $\chi^2$ is $\sim 50$ for seven degrees of freedom. As an additional test, we repeat the inference, this time not including image fluxes as a constraint (row two of Table \ref{tab:modposflux}). In this case, the fit is much better, with a $\chi^2$ of $\sim$1 for four degrees of freedom. The inferred model parameters in this case are consistent with previous studies which used IR and radio data but did not include flux information in their lens models
 \citep{Mao++98, Keeton++01, Bradac++05, Sluse++12b}.

\begin{figure*}
\centering
\includegraphics[scale=0.45]{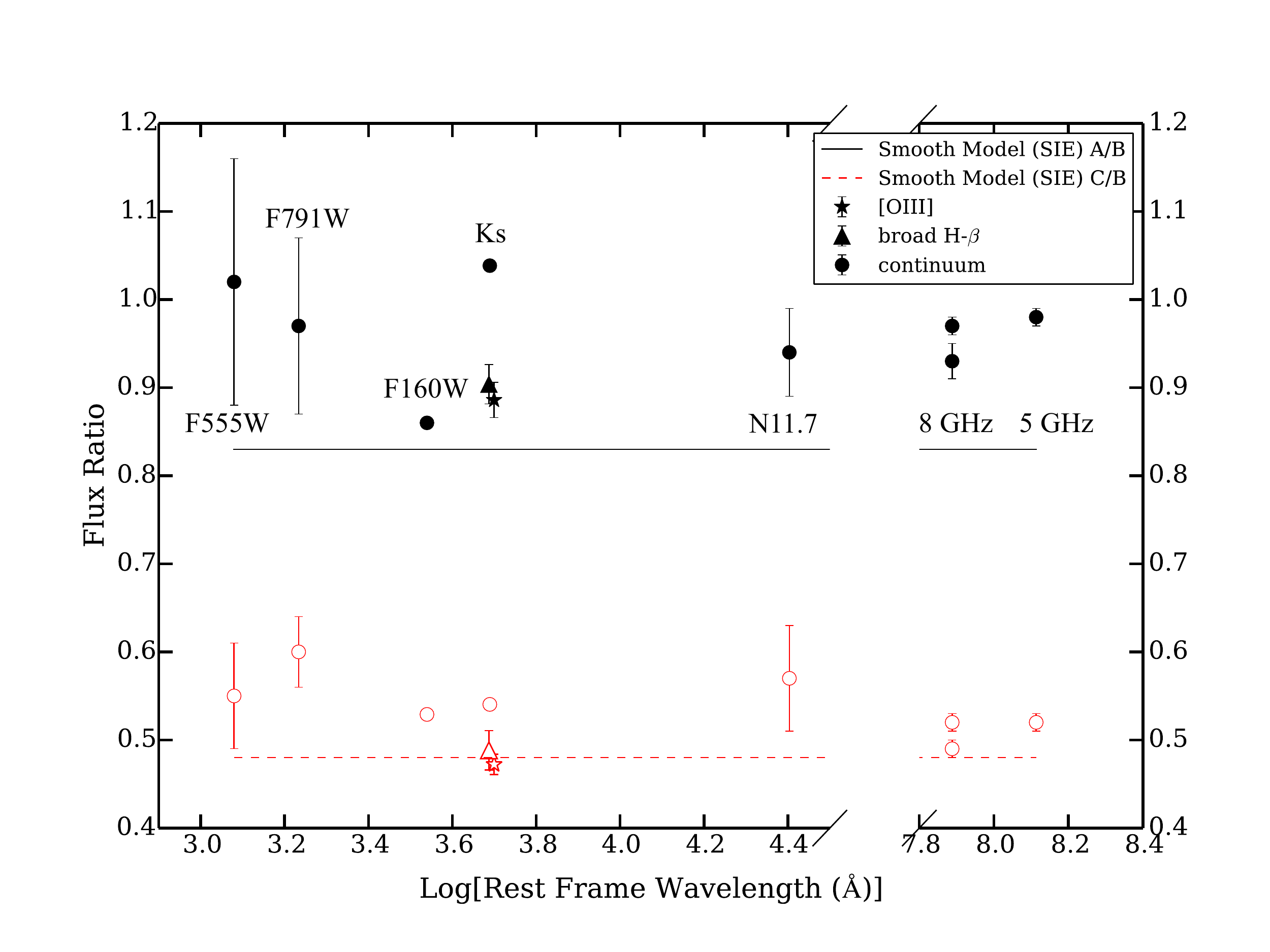}
\caption{ Comparison of measured ratios of images A B and C as a function of rest frame wavelength, with the narrow (star), broad (triangles) and continuum Ks measurement from this work. Continuum measurements are labelled with the observed filter for convenience.
We include HST measurements with WFPC2 F791W and F555W measurements from the CASTLES website, while the NICMOS F160W measurement is from \citet{Sluse++12b}. The mid-IR N11.7 measurement is from \citet{Chiba++05}. Radio measurements are from \citet{Patnaik++92,Patnaik++1999}. The solid and dashed lines represent the best-fit prediction to the [OIII] fluxes for the flux ratios between images A/B and C/B respectively, assuming one lens galaxy which is a singular isothermal ellipsoid with external shear. While the ratio between image B and C agrees well with the model, the ratio between A/B is significantly different from the smooth model prediction. \label{fig:fluxRatios} }
\end{figure*}

 The significant deviation between the observed [OIII] image flux and the smooth model prediction is consistent with previous observations of the system across a broad range of wavelengths, which also found that flux of image A is about 20\% brighter relative to image B than predicted by the single deflector model  \citep{Mao++98, Keeton++01, Bradac++05, Sluse++12b, Dobler++06}. Figure \ref{fig:fluxRatios} compares the [OIII] flux ratios between images A, B and C with measurements made at other wavelengths, and also relative to the smooth model prediction.

In Figure \ref{fig:fluxRatios}, we also show our measured flux ratios in the continuum and broad H$\beta$ emission. The broad H$\beta$ emission has flux ratios consistent with that of the narrow emission, while the continuum emission deviates more dramatically from the smooth model prediction. This is consistent with microlensing which can act in conjunction with millilensing by a subhalo, and would most strongly affect the smaller continuum source size.  Note that while the macromodel can be adjusted to provide a reasonable fit to the C/B flux ratios at other wavelengths, this is not possible in the case of the A/B flux ratios.


In order to improve the model fit, a natural solution is to add a perturber less massive than the central galaxy. This can alter image fluxes without significantly shifting image positions, because image fluxes depend on the second derivative of the lensing potential, while image positions depend on the first derivative. In the following subsection we explore the effects of adding a single perturbing mass to the lens system.

\subsection{Smooth model plus Perturber}
\label{sec:perturbed}

As can be seen in Figure \ref{fig:fluxRatios}, the ratio between the fluxes of images A and B deviates significantly from the smooth model prediction while the ratio between fluxes of image C and B does not. The simplest solution is that image A has been magnified by a nearby perturber \citep{Mao++98,Dobler++06,Bradac++05}, with the amount of magnification depending on the perturber position, mass scale and mass profile. 

We explore the simple case of adding a single perturber to the SIE plus external shear model. This is a useful way to understand how our astrometric and photometric precision allows us to constrain the presence of substructure, and to compare our results with previous works, which have also assumed a single perturbing subhalo to the smooth lens model. In reality, the system is likely to contain thousands of lower mass subhalos. However the majority of these subhalos will be too low in mass and too far from the lensed images to affect them. The degeneracies between perturber positions and mass make it difficult to distinguish between the effects of populations of perturbers and a single perturber added to the main lens system \citep{Fadely++12}. With a larger sample of lens systems, the statistical properties of the perturber population can be inferred more robustly, thus we will leave a more physical model with a realistic population of substructure to a future work in which we consider a larger sample of lensed quasar systems. 

For illustrative purposes we consider three mass profiles for the perturber, in order to explore how the mass profile affects the perturber lensing signal. Below we discuss each of the mass profiles in detail.

The simplest mass profile computationally is the singular isothermal sphere (SIS) given by Equation \ref{eq:sis}, with ellipticity set to zero. This profile is not physical as it diverges as $r$ goes to zero and extends outward indefinitely, however it provides a good approximation to the small scale density profiles of galaxies \citep[e.g.][]{Gav++07,Koopmans++09,Lagattuta++10}.

It is expected that a subhalo will undergo truncation by tidal forces from the main halo. This truncation can be modelled by a pseudo-Jaffe profile \citep[PJ, ][]{Munoz++01}, which has a density distribution:
\be
\rho(r) = \frac{\rho_o}{r^2}\frac{a^4}{r^2+a^2},
\ee
where $\rho_o$ is the same as in Equation \ref{eq:sis}. The truncation radius, $a$, is traditionally calculated by assuming that the perturber is exactly within the plane of the lens galaxy at the Einstein radius, $b_{host}$, so that $a = \sqrt{b_{\rm{sub}}b_{\rm host}}$ \citep{Metcalf++01}. In reality, tidal stripping depends on a variety of factors, including the orbit of the perturber around the host galaxy. This assumption for the tidal truncation radius assumes that the current projected position is also the pericentre of the satellite orbit. The SIS and PJ models bracket two extreme cases of no tidal stripping on the one hand, and maximum tidal stripping (in the absence of baryons) on the other.

The third mass profile we consider is the Navarro Frenk and White \citep[][NFW]{NFW++1996} profile. This mass profile is predicted by dark matter-only simulations, which may be a better match to the mass profile of subhalos than the SIS profiles. In fact the mass profile of subhalos may be even shallower than this at the centre \citep[e.g.][]{Walker++11, Wolf++12, Hayashi++12}, but we leave a more complete exploration of the effects of mass profile variations to a future work. The density distribution of the NFW halo is:
 \be
 \rho(r) = \frac{\rho_s}{(r/r_s)(1+r/r_s)^2}
\ee
Where $r_s$ and $\rho_s$ are the scale radius and the density at the scale radius respectively. 
To reduce the number of free parameters, we apply the mass-concentration relation predicted by \citet{Maccio++08}, assuming a WMAP5 cosmology \citep{Dunkley++09}. This is a large extrapolation at dwarf galaxy scales, but allows us to make a useful one-parameter comparison with the steeper mass profiles.

 As a first step, we again use {\tt gravlens} to find the best-fit model to the narrow-line fluxes and positions for the macromodel parameters (host and external shear) as well as the perturber mass and position in the case of each of the three perturber mass profiles. As with the SIE only model, we then use {\tt emcee hammer} to infer the model parameters, allowing both the macromodel and the substructure parameters to vary, assuming a uniform prior on the perturber position and log-uniform prior on the perturber mass. For each proposed perturber mass scale, position and set of host properties, we use {\tt gravlens} in order to optimise the source position and compute the $\chi^2$ for the resulting image positions and magnifications.
 
 As we will demonstrate below, despite the degeneracies, the possible perturber positions are restricted by the proximity of image B to image A, as well as by the critical curve of the main lens.  Furthermore, the astrometric precision of the measurement prevents a subhalo large enough to cause significant astrometric deviations from the smooth model. These effects can all be accounted for in the {\tt gravlens} code. We also place lower limits on the mass of the perturber based on the finite size of the source, which we describe below.
  
\subsection{Finite Source Effects}

The finite nature of the source places lower limit to the perturber mass that can cause the observed magnification \citep{D+K06}.
This can qualitatively be understood by the fact that observed magnification is the convolution of the magnification pattern with the source surface brightness distribution. The larger the source the more small-scale features in the magnification due to low-mass substructure will be smeared out. In the following subsection we discuss how we infer the narrow-line emission source size and use this to place lower limits on the perturber mass for the three different perturber mass profiles.

The observed images in each exposure are a convolution of the PSF in that exposure which we modelled as the sum of two concentric Gaussians in Section \ref{sec:imExtract}, with the source which we assumed to be a point source. This was a valid assumption given that the flux in the images between 4500-5100 \AA~ is dominated by continuum and broad H$-\beta$ emission, which are point-like ($\mu$as) relative to the resolution of the telescope. Narrow-line emission, on the other hand, is seen to extend out to hundreds or even thousands of parsecs for the most luminous Seyferts at low redshift \citep[e.g.][]{Bennert++02}, although the dominant contribution to the narrow-line luminosity comes from the central tens of parsecs \citep{Muller++11}.

In order to determine the size of the lensed narrow-line source, we adjust the analysis described in Section \ref{sec:imExtract}, this time modelling the flux as being emitted from a Gaussian light distribution. This Gaussian source, when convolved with the PSF Gaussian yields observed image widths given by 
\be
\sigma_{eff} = \sqrt{\sigma_{PSF}^2+ (\mu^{1/2} \sigma_{source})^2} ,
\ee

where $\sigma_{PSF}^2$ is the standard deviation of the seeing-halo, and $\mu$ is the best fit magnification given by {\tt gravlens} at each of the image positions. 

We apply this analysis to a narrow-line image, created by subtracting a variance weighted mean image of the continuum region between rest-frame 5070-5140 \AA ~ from a variance weighted mean image which contains [OIII] plus continuum between rest frame 4980-5030 ~\AA. We infer an intrinsic source size of $\sim 15$ mas. This analysis assumes that the diffraction core of the AO corrected PSF had a fixed size of 6.5 mas, while in reality the size of the diffraction core depends on the AO performance and weather conditions, and can vary from night to night. To test this, we also infer the source size in an image containing light from only the quasar continuum, which we expect to be unresolved. Instead, we find that the continuum image has an intrinsic source size of $\sim$ 10 mas, implying that the diffraction core of the PSF is not as narrow in reality as in our model. However, the [OIII] emission is definitely larger and thus resolved. Subtracting the two in quadrature we estimate the intrinsic [OIII] size to be roughly $\sim10$ mas. This corresponds to approximately 60 pc at the redshift of the quasar.

Based on the size-luminosity relation for type-I AGN measured in the Local Universe by \citet{Bennert++02}, we would have expected the full narrow-line region in B1422+231 to extend to several hundreds of pc. However, we are likely to only be sensitive to the highest surface brightness parts of the narrow line regions because of surface brightness sensitivity and cosmological dimming. So it is expected that our measurement be smaller than in the local universe.
Furthermore, we do not consider lensing shear distortions which are expected to cause the images to be elongated tangentially to the radial direction, rather than simply enlarged isotropically as our analysis assumes. Thus this simple calculation should be considered a lower limit to the true narrow line emission region size, which corresponds to a conservative lower limit to the perturber mass. In other words, a larger narrow line region would result in a more narrow posterior distribution function of the perturber mass.
 
\citet{Dobler++06}, computed the relationship between the source size and the minimum Einstein radius of an SIS perturber that can cause the observed magnification of image A, given the convergence and shear of the system B1422+231. They found that an SIS perturber must have mass within the Einstein of $b>0.056 a$, where $b$ is the Einstein radius of the SIS perturber, and $a$ is the unlensed source size. This corresponds to a lower mass limit of $\sim10^{6.5}$ and $\sim10^6$ M$_\odot$ within 600 pc in the case of the SIS and PJ perturbers respectively, assuming the perturbers are in the plane of the lens galaxy\footnote{The interpretation of the enclosed mass given a fixed Einstein radius and density profile depends on the perturber redshift.}

The mass limit on the NFW perturber cannot be computed analytically. We approximate this limit by computing the average magnification in a grid of points with a size of the magnified image, for increasing perturber masses. The limit is then given approximately by the minimum perturber mass that can achieve the observed magnification in a region the size of the magnified image. This is an approximate result, as it assumes the source is magnified isotropically rather than distorted as it actually is. In the case of the SIS perturber, for instance, with this method we find that the perturber must have $b >0.04 a$, which is slightly lower than the true value computed by \citet{Dobler++06}, but is good enough for an order of magnitude estimate of the minimum perturber mass. In the case of an NFW perturber with mass concentration relation given by \citet{Maccio++08}, the perturber must have a scale radius which is approximately five times larger than the unlensed source size, which corresponds to a mass limit of $\sim 10^{6.5}$M$_\odot$ within 600 pc.

\section{Results}
\label{sec:results}
The SIE host plus perturber lens models show significant improvement relative to the SIE-only model for all three perturber mass profiles, with best fit $\chi^2$ of one or smaller for four degrees of freedom. The model prediction for the image positions and fluxes are listed in Table \ref{tab:modposflux}, for the case of an SIE host with an SIS perturber. 
The model parameters which determine the properties of the main SIE lens and external shear field are consistent for the SIE plus perturber model and the  SIE only model when fluxes were not included as a model constraint. This illustrates how the perturber can alter image fluxes without significantly affecting image positions.


\begin{table*}
\centering
\scriptsize\begin{tabular}{lllllllll|l}
\hline \hline
Model & host b & $\epsilon$ & host PA & $\gamma$& $\theta_{\gamma}$ &$\log_{10}[\rm{b}_{\rm{SIS}}]$ & dRA$_{\rm{SIS}}$ & dDec$_{\rm{SIS}}$  &$\chi^2$/DOF    \\
\hline 


SIE  (fluxes)&    0.746$\pm$0.007   & 0.36$\pm0.03$ & -62$\pm1$&0.16$\pm0.01$ & -47$\pm2$&- & - &- & 7    \\
SIE (no fluxes) &    0.771$^{+0.007}_{-0.009}$&  0.16$\pm0.08$& -57$^\pm5$ & 0.22$\pm0.03$ & -54$\pm1$&- & - &-& -  \\
SIE + SIS (fluxes)& 0.765$\pm0.009$ & 0.17$^\pm0.07$& -58$\pm5$ & 0.22$\pm0.02$ &  -54$^{+2}_{-1 }$ &-1.9$^{+0.6}_{-0.7}$ &0.5$^{+0.1}_{-0.2 }$& 0.4$\pm0.2$ & 0.1 \\
\hline \hline
\end{tabular}
\caption{Posterior median and 68\% confidence intervals for lens model parameters, and maximum likelihood $\chi^2$ per degree of freedom for the two cases in which the full data set is used. In the first two cases, the lens was modelled as an SIE lens with external shear, first using image fluxes and positions as well as the galaxy position to constrain the model, in the second case using only the image and galaxy positions to constrain the model, for comparison with previous works. In the third row, the lens is modelled as an SIE in external shear with a perturbing subhalo with an SIS mass profile, using image fluxes and positions, and the galaxy position as constraints. Results are inferred based on the [OIII] image fluxes and positions. Parameters are defined in Section \ref{sec:smooth}. Position angle and shear are in units of degrees East of North while Einstein radii are in units of arcseconds. 
\label{tab:modelPars}}
\end{table*}

\begin{table*}
\scriptsize\begin{tabular}{llll|lll}
\hline \hline
Image & dRa$_1$ & dDec$_1$ & Flux$_1$ & dRa$_2$ & dDec$_2$ & Flux$_2$\\
\hline 

A & 0.375    $\pm 0.003$  & 0.310$\pm 0.004$  & 0.83  $\pm$ 0.01       & 0.376 $\pm$ 0.004    & 0.327  $\pm$ 0.005       &  0.88  $\pm$ 0.02 \\ 
B &  0.002    $\pm 0.003$  & 0.014 $\pm 0.004$ & 1.00                              & 0.000$\pm 0.004$     & 0.000 $\pm 0.005 $       &  1.00         \\ 
C & - 0.346  $\pm 0.005$  &-0.743 $\pm 0.003$ & 0.484 $\pm$ 0.006       & -0.339 $\pm 0.005$   & -0.739 $\pm 0.004$       &  0.471  $\pm$ 0.007 \\
\hline \hline
\end{tabular}
\caption{Posterior median and 68\% confidence interval for model predicted image positions in units of arcseconds and fluxes in the case of a single smooth lens galaxy (1) and a lens galaxy plus perturber with SIS mass profile (2). Results for the other two smooth plus perturber mass profiles are similar. \label{tab:modposflux}}
\end{table*}

In Table \ref{tab:modelPars} we list the median and one sigma confidence intervals for the lens model parameters in the case of an SIE host galaxy and an SIS perturber. The host model parameters are consistent with those in the smooth model case, which is expected given the fact that the perturbing halo is a relatively small addition to the mass of the system.

It is useful to quantify the significance of the perturbation to the smooth model. Here we discuss three different model testing methods. First, the inferred distribution for the perturber position and mass is informative given that the SIE plus perturber models allows for solutions in which the perturber does not contribute significantly to the lensing, for instance if it were low mass and far from the lensed images. In this sense the more complex SIE plus perturber model also includes the simpler possibility of the SIE-only model. We find that in 1.8 million iterations of the MCMC, the perturber is always located such that it is contributing significantly to the magnification of image A relative to the smooth model at that location.

In addition, the maximum likelihood $\chi^2$ per degree of freedom for the SIE-only model is 7 and the for SIE plus SIS perturber model it is 0.1. Results are similar for the other two perturber mass profiles. 
An alternative way to compare models is via the Akaike Information Criterion, \citep{Akaike++74} modified to take into account the relatively large number of model parameters relative to data in this case \citep[AICc][]{Hurvich++89}. This is useful when trying to determine whether the addition of new parameters to a model improves the fit relative to the `true' underlying model \citep{Kelly++14}.

The AICc is calculated as:
\be
AICc = 2 k - 2\log p_{\rm{ml}} + \frac{2 k(k+1)}{n-k-1}
\ee
Where here $k$ is the number of model parameters, $p_{\rm{ml}}$ is the maximum likelihood fit to the data given a model choice, and $n$ is the number of independent data points used to constrain the model. The AICc is similar to the $\chi^2$ with an additional penalty for extra model parameters. 
Given maximum $\log p$ values of the SIE only and SIE plus SIS models, the AICc values are 60 and 50. This indicates that the SIE plus SIS model is of order 100 times more likely than the SIE only model.

\begin{figure}
\centering
\includegraphics[scale=0.3, trim = 15 0 0 40, clip = true]{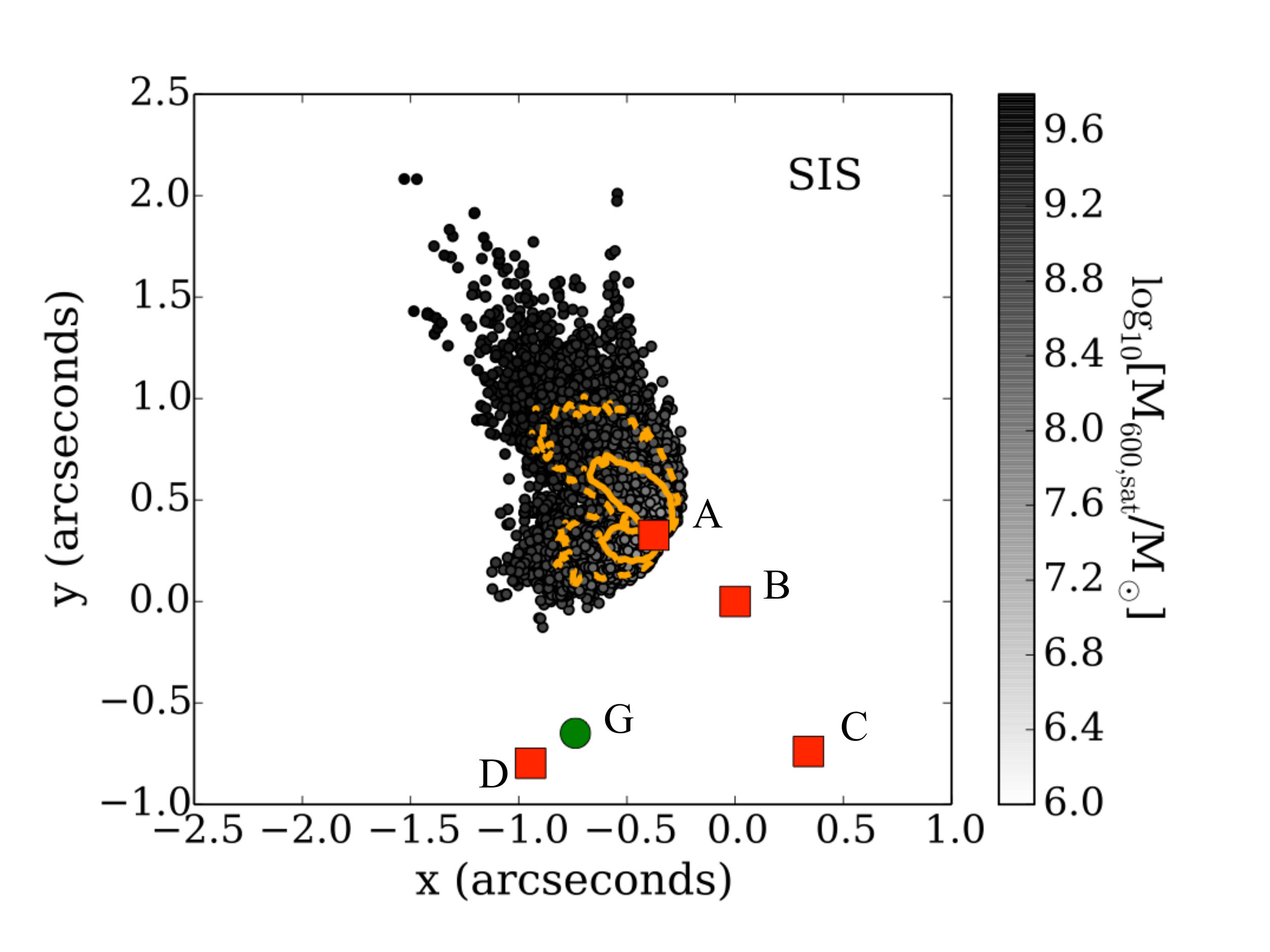}\\
\includegraphics[scale=0.2,trim = 15 0 0 40, clip = true]{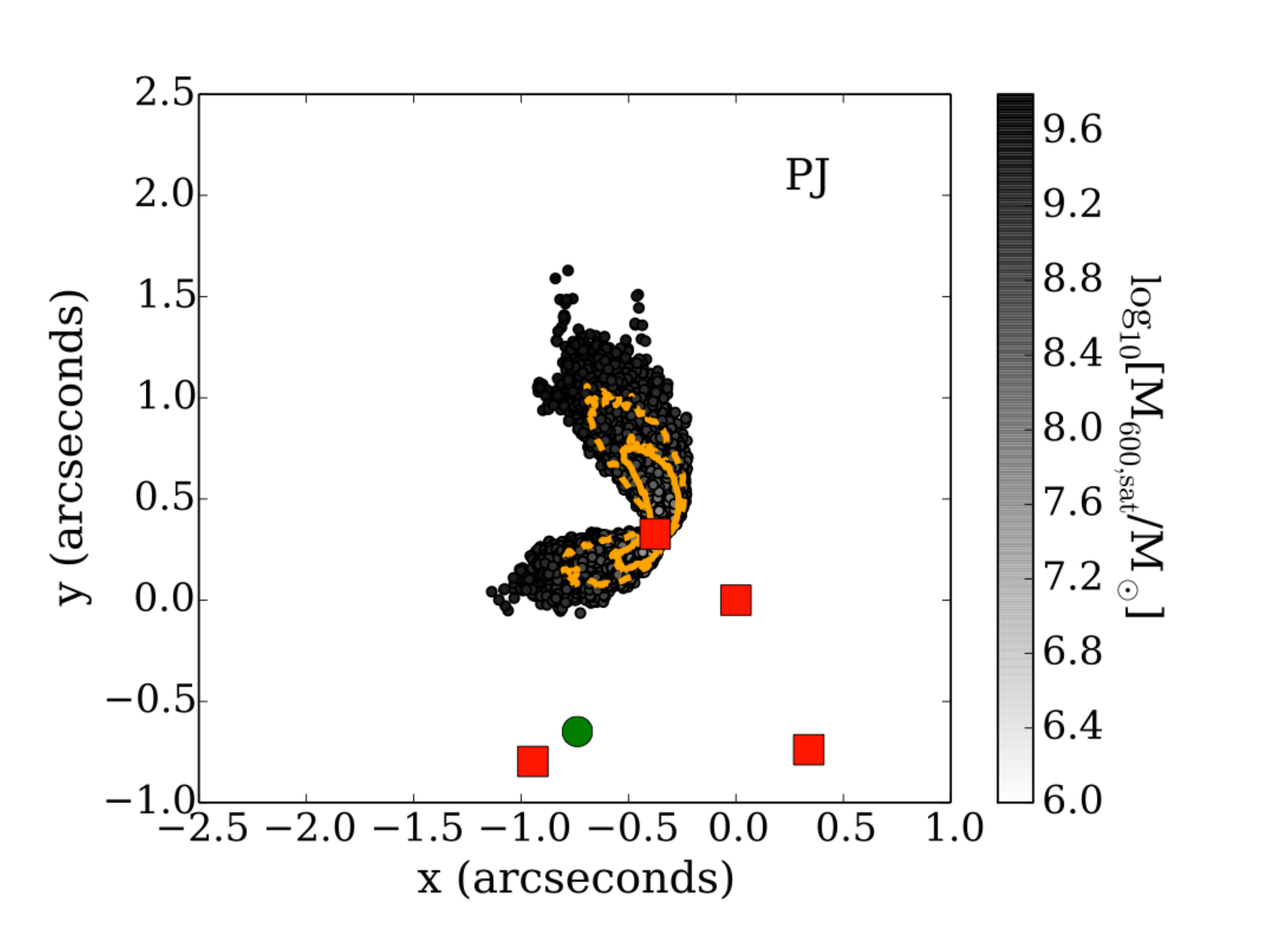}\\
\includegraphics[scale=0.2,trim = 15 10 0 40, clip = true]{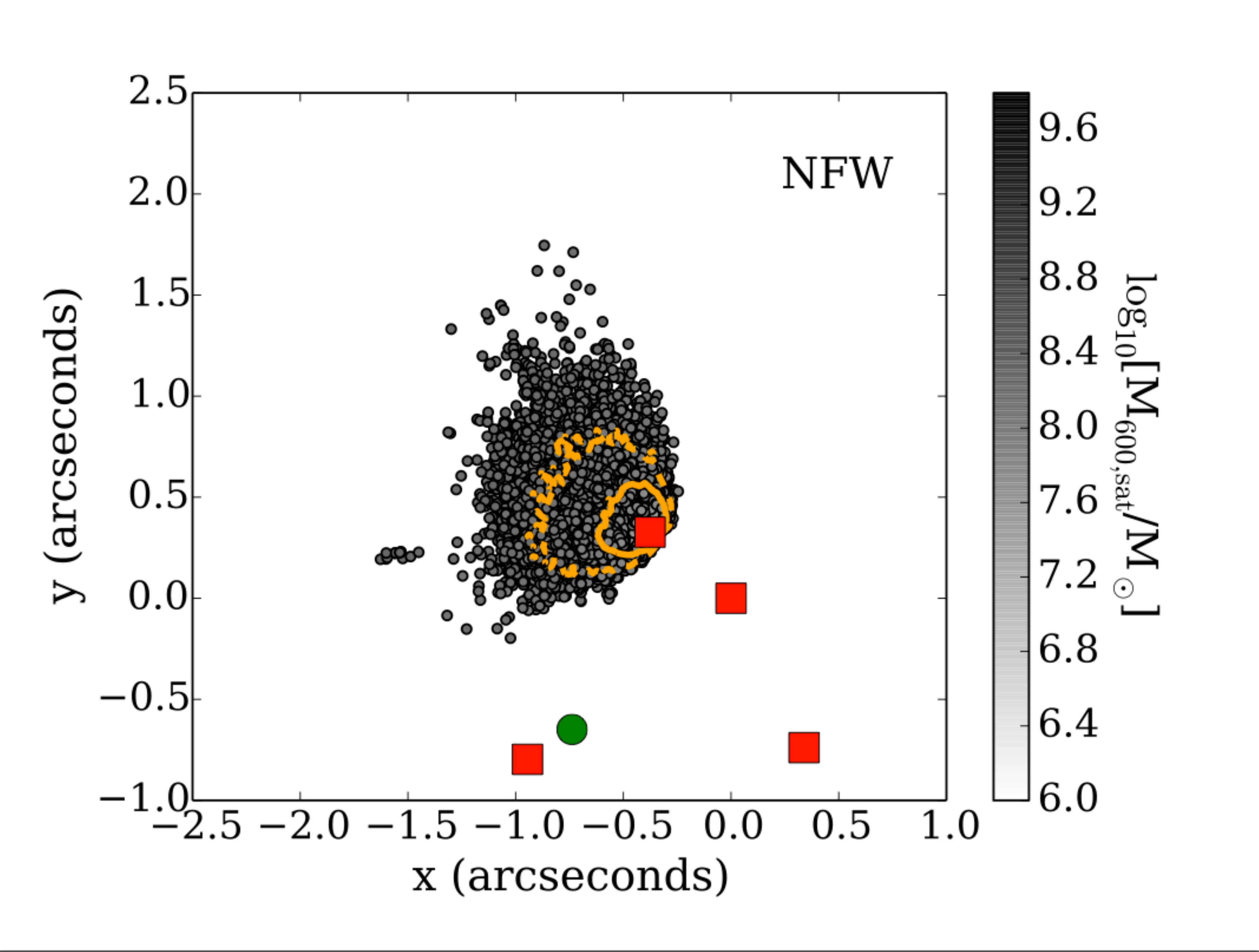}
\caption{Posterior probability distributions of the perturber position relative to the lensed images shown as red squares, and lens galaxy shown as a green circle, for a single SIS, PJ, and NFW perturber from top to bottom. The grey scale represents the perturber mass within 600 pc assuming the perturber is in the plane of the lens galaxy, and solid and dashed contours represent the 68 and 95\% confidence contours respectively relative to the most likely position.}
\label{fig:prmass}
\end{figure}

While all three perturber mass profiles provide equally good fits to the observations, the posterior probability distributions for the perturber's position and mass vary significantly as can be seen in Figure \ref{fig:prmass}. There is similar qualitative behaviour in the sense that the further a perturber is from image A, the more massive it must be to achieve the same lensing effect. Furthermore, in all three cases, the mass range is limited naturally by the astrometric precision of the measurement. If the perturber becomes too massive it causes significant astrometric perturbations to the whole system, rather than just affecting the flux of image A.  However, the posterior probability distribution for the SIS perturber position is much broader than in the case of the PJ profile. This is due to the fact that the SIS mass profile extends outward indefinitely, so that it has a fundamentally different lensing effect than the PJ perturber outside of the truncation radius of the PJ perturber. We discuss this in more detail in the Appendix.  Another interesting feature is that the shallower NFW profile is not restricted by the position of the lens critical curves in the way the SIS and PJ perturbers are. This shallower profile  integrates to a larger total mass for fixed aperture mass. Given that the lensing effect is determined by the aperture rather than total mass,  this implies that to achieve the same magnification as the SIS or PJ perturbers, the total mass of the NFW perturber must be higher. This limits the 95\% and 68\% position contours for the NFW profile to be closer to image A than the other two profiles.

\begin{figure*}
\centering
\includegraphics[scale=0.55, trim = 0 0 0 0, clip = true]{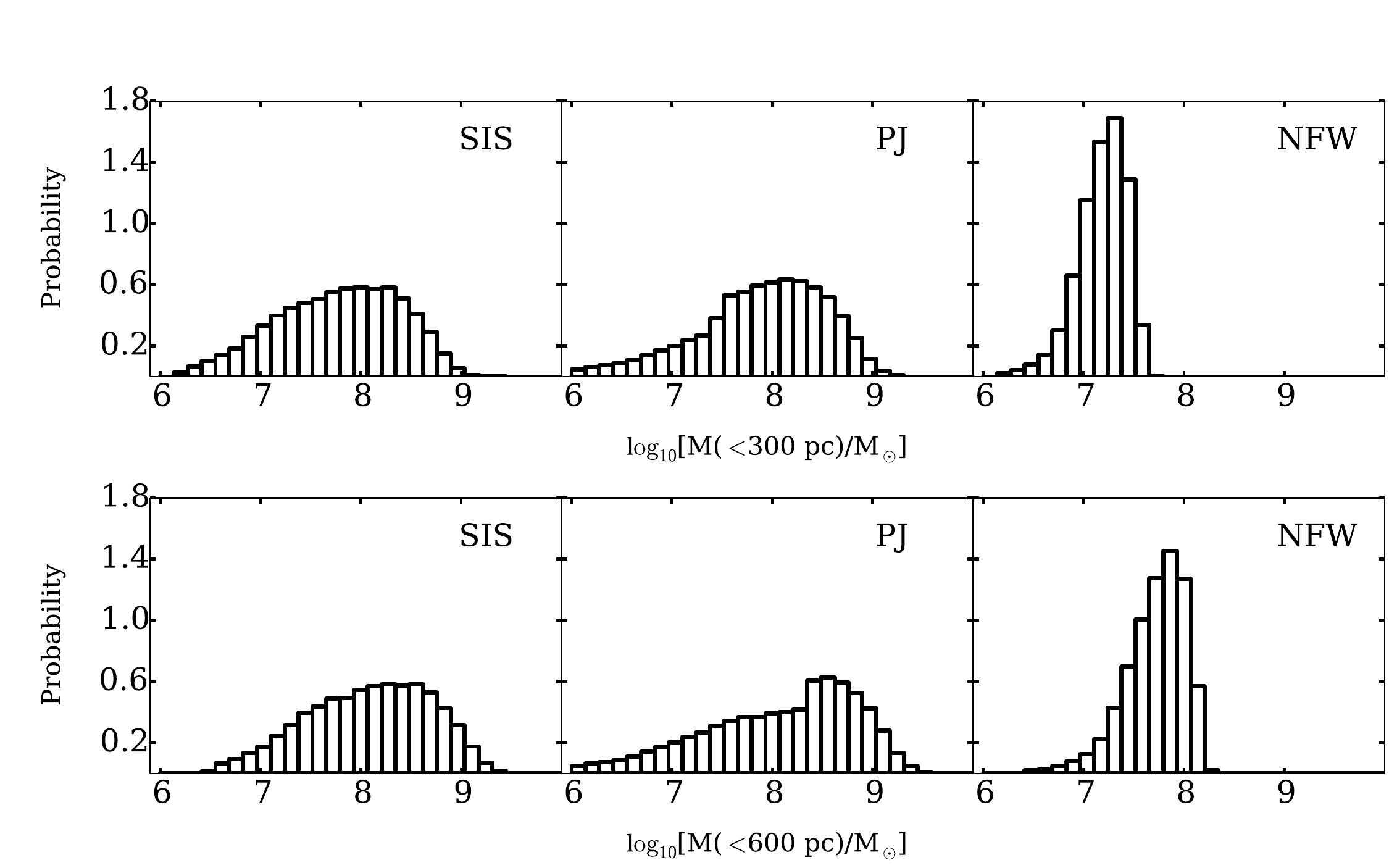}
\caption{Marginalised posterior probability distribution of the perturber mass within 300 pc (upper row) and 600 pc (lower row) assuming an SIS, PJ and NFW mass profile respectively from left to right.}
\label{fig:pmass}
\end{figure*}

In Figure \ref{fig:pmass}, we plot the marginalised posterior probability distribution for the perturber mass within 600 parsecs, assuming that the perturber is located in the plane of the lens galaxy. We note that for PJ perturbers with masses lower than $\sim10^{8.5}$, the truncation radius falls inside of 600 pc. The 68\% confidence intervals for the logarithm of the perturber mass within 600 pc are  8.2$^{+0.6}_{-0.8}$, 8.2$^{+0.6}_{-1}$ and 7.6$^{+0.3}_{-0.3}$ $\log_{10}$[M$_{\rm{sub}}$/M$_\odot$], for a singular isothermal sphere, PJ, and NFW mass profile respectively. The PJ perturber can have lower masses relative to the SIS mass profile due to the truncation radius which simulates tidal stripping, and falls within 600 pc for PJ perturbers with masses less than $\sim10^{8.5}$M$_\odot$ within 600 pc. The NFW profile is restricted to a small range of masses due to its shallow density profile which gives it an effectively weaker lensing signal.  For ease of comparison with Local Group studies, the mass within 300 parsecs is shown in the upper panel of Figure \ref{fig:pmass}, and has a 68\% confidence interval of 7.8$^{+0.6}_{-0.7}$,8.0$^{+0.6}_{-0.8}$, 7.2$^{+0.2}_{-0.2}$ $\log_{10}$[M$_{\rm{sub}}$/M$_\odot$] for the three mass profiles.

Given the significantly different lensing effects in the three cases, it will be possible with a larger sample of lenses to learn about both the typical mass profile as well as the mass function of the perturbers. 
 
\section{Discussion}
\label{sec:discussion}
 
 Our primary goal in this work is to demonstrate that strongly lensed narrow-line emission provides an alternative to radio emission in lensed quasars for detecting substructure, yielding lens model constraints with consistent results and comparable accuracy. In this section we discuss previous results for this system, how the analysis can be improved by considering additional systematic uncertainties, as well as future prospects for the method

\subsection{Comparison with previous work} 
B1422+231 is a bright, radio loud system which enabled one of the earliest detections of dark substructure at cosmological distances \citep[e.g.][]{Mao++98,Keeton++01, Bradac++05}. These studies found that the observed radio fluxes could be best explained by a perturbing substructure with characteristic Einstein radius of a few mas. Although there was some question about how macromodel assumptions might affect the inference of substructure \citep{Kawano++04}, \citet{Kochanek++04} demonstrated that smooth macromodels that are consistent with observations of galaxies using weak lensing and galaxy morphology do not provide a good fit to the observed image positions and flux ratios. Our analysis based on lensed narrow-line emission yields consistent results to these studies.
 
\subsection{Limitations of our analysis}
We have explored how our data can be used to constrain a mass model with a main lens in 
external shear and one perturbing subhalo. As we discussed in Section \ref{sec:perturbed}, 
this model makes several important simplifying assumptions. 

First, for simplicity we adopt a macro-model composed by a singular
isothermal ellipsoid with external shear. This is in general found to
be a good model for early-type galaxies both in terms of radial
profile \cite{Koo++06,Koo++09} and angular structure
\citep{Yoo++05,Yoo++06}. However, allowing for more flexibility in the
model could inflate the uncertainties on the inferred substructure, or
reduce its significance, even though in the case of B1422+231 it is
hard to find physically plausible models that fit the data without a
localized substructure \citep{E+W03}.

Second, as with previous works, we explore the effects of a single
perturbing subhalo \citep{MacLeod++09,McKean++07,MacLeod++13}. When
studying the effects of subhalos on lensed compact image fluxes and
positions, full populations of subhalos are computationally expensive
and not well constrained by a single lensing system
\citep{Fadely++12}. In the future, when large samples will be available, it will be important to carry out a systematic analysis of the entire population of subhalos using hierarchical modelling to infer at once the mass function and spatial distribution of the subhalos.

Third, similarly to other works of this type, we have assumed that the
perturbing substructure is in the plane of lens galaxy in order to
estimate the perturber mass \citep[e.g.][]{Dalal++02,
MacLeod++09,Fadely++12,Vegetti++10a,Vegetti++12,Vegetti++14}.
\citet{Xu++12} showed that line of sight structure may contribute
significantly to image magnifications, although the observed frequency of 
radio flux ratio anomalies can be explained with in 
situ perturbers predicted by CDM \citep{Xu++14, Metcalf++12}.
This has important implications for the inferred
subhalo mass function as all masses are inferred assuming the redshift
of the perturber is known. The lensing effect varies with mass if the
perturber is not within the plane of the lens galaxy
\citep{Xu++12,McCully++2014}.

We have also demonstrated the effects of varying the assumed mass profile of the perturbing subhalo, which significantly change the posterior probability distribution for the perturber mass and position. It would be interesting in a future work to consider a range of subhalo mass profiles which are drawn directly from simulations, so that differential tidal stripping as a function of three dimensional distance from the lens halo centre, and scatter in the subhalo concentration mass relation can be taken into account for instance.

\subsection{Future Prospects}
Many more quasars have significant narrow-line emission than radio emission, thus this method can be extended to study substructure in a larger sample of systems. There is significant future potential for this method as ongoing and planned optical surveys such as DES, LSST, PANSTARRS and GAIA are expected to find thousands more lensed quasars based on their optical rather than radio properties \citep{Oguri++10}. The next generation of adaptive optics systems (NGAO), \citep{Max++08}, and telescopes such as the Thirty Meter Telescope and the European Extremely Large Telescope will make it possible to measure narrow-line lensing rapidly and with high precision for those large samples of objects. In the nearer future, upgrades to the Keck AO system will enable PSF telemetry which will thereby make it possible to obtain better constraints on the source size, and thus place stronger lower limits on perturber masses.


Significant information is also contained in deep imaging of the lensed systems, which can sometimes reveal luminous substructure \citep{McKean++07, MacLeod++09}. If a luminous satellite is detected, it can be used to break the degeneracy between the subhalo mass and position, as well as providing a measurement of the subhalo mass to light ratio.

\section{Summary}
\label{sec:summary}

We used OSIRIS at Keck with adaptive optics in order to obtain spatially resolved spectroscopy of the gravitational lens B1422+231. We developed a new pipeline which enabled us to robustly extract lensed image spectra, by first inferring the PSF and image positions in each exposure. 
Using this information, we detected a significant deviation from a smooth gravitational lens model, which we used to infer the presence of a perturbing low mass subhalo. Our main results are summarised below:

\begin{enumerate}
\item Using our data reduction pipeline, we were able to measure image positions to 5 mas accuracy and integrated line flux ratios with $\sim$3\% uncertainties. 

\item The narrow-line flux ratios are consistent with radio measurements, and deviate significantly from the smooth model predictions. The broad-line flux ratios are consistent with the narrow line measurements.  Our measurement of the continuum is significantly offset, showing a much stronger deviation from the smooth model, possibly indicative of microlensing in addition to millilensing. 

\item 
Based on the assumption that the deviation of observed narrow-line fluxes from the smooth model is due to a single perturbing subhalo, we infer the mass and position of the perturbing subhalo for three different mass profiles.
The mass profile significantly affects possible perturber positions and masses, with an NFW subhalo restricted to the smallest region of parameter space by its shallower mass profile which has a relatively weaker lensing signal. The 68\% confidence intervals for the logarithmic perturber mass within 600 pc are:  8.2$^{+0.6}_{-0.8}$, 8.2$^{+0.6}_{-1}$ and 7.6$\pm 0.3$ $\log_{10}$[M$_{\rm{sub}}$/M$_\odot$], for a singular isothermal sphere, pseudo-Jaffe, and NFW mass profile respectively.  The mass within 300 pc is 7.8$^{+0.6}_{-0.7}$ and 8.0$^{+0.6}_{-0.8}$, 7.2$^{+0.2}_{-0.2}$$\log_{10}$[M$_{\rm{sub}}$/M$_\odot$]

\end{enumerate}

\section*{Acknowledgments}
We thank B. Kelly, M. Bradac, S. Vegetti, D. Sluse, C. Keeton, S. Mao, G. Dobler, P. J. Marshall, B. Brewer, V. Bennert, A. Pancoast, S. Suyu and C. Kochanek for useful comments and discussion. We also thank support astronomer S. Dahm and observing assistant T. Stickel. 
AMN and TT acknowledge support
by the NSF through CAREER award NSF-0642621, and by the Packard
Foundation through a Packard Fellowship. AMN acknowledges support from a UCSB Dean's Fellowship.
 
\bibliographystyle{apj}
\bibliography{references}

\begin{thebibliography}{94}
\expandafter\ifx\csname natexlab\endcsname\relax\def\natexlab#1{#1}\fi

\bibitem[{{Akaike}(1974)}]{Akaike++74}
{Akaike}, H. 1974, Automatic Control, IEEE Transactions on, 19, 716

\bibitem[{{Amara} {et~al.}(2006){Amara}, {Metcalf}, {Cox}, \&
  {Ostriker}}]{Ama++06}
{Amara}, A., {Metcalf}, R.~B., {Cox}, T.~J., \& {Ostriker}, J.~P. 2006, \mnras,
  367, 1367

\bibitem[{{Bennert} {et~al.}(2002){Bennert}, {Falcke}, {Schulz}, {Wilson}, \&
  {Wills}}]{Bennert++02}
{Bennert}, N., {Falcke}, H., {Schulz}, H., {Wilson}, A.~S., \& {Wills}, B.~J.
  2002, \apjl, 574, L105

\bibitem[{{Benson} {et~al.}(2002){Benson}, {Frenk}, {Lacey}, {Baugh}, \&
  {Cole}}]{Benson++2002}
{Benson}, A.~J., {Frenk}, C.~S., {Lacey}, C.~G., {Baugh}, C.~M., \& {Cole}, S.
  2002, \mnras, 333, 177

\bibitem[{{Boylan-Kolchin} {et~al.}(2011){Boylan-Kolchin}, {Bullock}, \&
  {Kaplinghat}}]{Boylan-Kolchin++11}
{Boylan-Kolchin}, M., {Bullock}, J.~S., \& {Kaplinghat}, M. 2011, \mnras, 415,
  L40

\bibitem[{{Boylan-Kolchin} {et~al.}(2012){Boylan-Kolchin}, {Bullock}, \&
  {Kaplinghat}}]{Boylan-Kolchin++12}
---. 2012, \mnras, 422, 1203

\bibitem[{{Brada{\v c}} {et~al.}(2002){Brada{\v c}}, {Schneider}, {Steinmetz},
  {Lombardi}, {King}, \& {Porcas}}]{Bradac++05}
{Brada{\v c}}, M., {Schneider}, P., {Steinmetz}, M., {Lombardi}, M., {King},
  L.~J., \& {Porcas}, R. 2002, \aap, 388, 373

\bibitem[{{Bullock} {et~al.}(2000){Bullock}, {Kravtsov}, \&
  {Weinberg}}]{Bullock++2000}
{Bullock}, J.~S., {Kravtsov}, A.~V., \& {Weinberg}, D.~H. 2000, \apj, 539, 517

\bibitem[{{Carlberg} {et~al.}(2012){Carlberg}, {Grillmair}, \&
  {Hetherington}}]{Carlberg++12}
{Carlberg}, R.~G., {Grillmair}, C.~J., \& {Hetherington}, N. 2012, \apj, 760,
  75

\bibitem[{{Chakrabarti} \& {Blitz}(2009)}]{Chakrabarti++09}
{Chakrabarti}, S., \& {Blitz}, L. 2009, \mnras, 399, L118

\bibitem[{{Chiba} {et~al.}(2005){Chiba}, {Minezaki}, {Kashikawa}, {Kataza}, \&
  {Inoue}}]{Chiba++05}
{Chiba}, M., {Minezaki}, T., {Kashikawa}, N., {Kataza}, H., \& {Inoue}, K.~T.
  2005, \apj, 627, 53

\bibitem[{{Dalal} \& {Kochanek}(2002)}]{Dalal++02}
{Dalal}, N., \& {Kochanek}, C.~S. 2002, \apj, 572, 25

\bibitem[{{Dobke} \& {King}(2006)}]{D+K06}
{Dobke}, B.~M., \& {King}, L.~J. 2006, \aap, 460, 647

\bibitem[{{Dobler} \& {Keeton}(2006)}]{Dobler++06}
{Dobler}, G., \& {Keeton}, C.~R. 2006, \mnras, 365, 1243

\bibitem[{{Dunkley} {et~al.}(2009){Dunkley}, {Spergel}, {Komatsu}, {Hinshaw},
  {Larson}, {Nolta}, {Odegard}, {Page}, {Bennett}, {Gold}, {Hill}, {Jarosik},
  {Weiland}, {Halpern}, {Kogut}, {Limon}, {Meyer}, {Tucker}, {Wollack}, \&
  {Wright}}]{Dunkley++09}
{Dunkley}, J., {Spergel}, D.~N., {Komatsu}, E., {Hinshaw}, G., {Larson}, D.,
  {Nolta}, M.~R., {Odegard}, N., {Page}, L., {Bennett}, C.~L., {Gold}, B.,
  {Hill}, R.~S., {Jarosik}, N., {Weiland}, J.~L., {Halpern}, M., {Kogut}, A.,
  {Limon}, M., {Meyer}, S.~S., {Tucker}, G.~S., {Wollack}, E., \& {Wright},
  E.~L. 2009, \apj, 701, 1804

\bibitem[{{Evans} \& {Witt}(2003)}]{E+W03}
{Evans}, N.~W., \& {Witt}, H.~J. 2003, \mnras, 345, 1351

\bibitem[{{Fadely} \& {Keeton}(2012)}]{Fadely++12}
{Fadely}, R., \& {Keeton}, C.~R. 2012, \mnras, 419, 936

\bibitem[{{Foreman-Mackey} {et~al.}(2013){Foreman-Mackey}, {Hogg}, {Lang}, \&
  {Goodman}}]{Mackey++13}
{Foreman-Mackey}, D., {Hogg}, D.~W., {Lang}, D., \& {Goodman}, J. 2013, \pasp,
  125, 306

\bibitem[{{Gavazzi} {et~al.}(2007){Gavazzi}, {Treu}, {Rhodes}, {Koopmans},
  {Bolton}, {Burles}, {Massey}, \& {Moustakas}}]{Gav++07}
{Gavazzi}, R., {Treu}, T., {Rhodes}, J.~D., {Koopmans}, L.~V.~E., {Bolton},
  A.~S., {Burles}, S., {Massey}, R.~J., \& {Moustakas}, L.~A. 2007, \apj, 667,
  176

\bibitem[{{Guerras} {et~al.}(2013){Guerras}, {Mediavilla}, {Jimenez-Vicente},
  {Kochanek}, {Mu{\~n}oz}, {Falco}, \& {Motta}}]{Guerras++13}
{Guerras}, E., {Mediavilla}, E., {Jimenez-Vicente}, J., {Kochanek}, C.~S.,
  {Mu{\~n}oz}, J.~A., {Falco}, E., \& {Motta}, V. 2013, \apj, 764, 160

\bibitem[{{Guo} {et~al.}(2011){Guo}, {Cole}, {Eke}, \& {Frenk}}]{Guo++11}
{Guo}, Q., {Cole}, S., {Eke}, V., \& {Frenk}, C. 2011, \mnras, 1278

\bibitem[{{Hayashi} \& {Chiba}(2012)}]{Hayashi++12}
{Hayashi}, K., \& {Chiba}, M. 2012, \apj, 755, 145

\bibitem[{{Hezaveh} {et~al.}(2013){Hezaveh}, {Dalal}, {Holder}, {Kuhlen},
  {Marrone}, {Murray}, \& {Vieira}}]{Hezaveh++13}
{Hezaveh}, Y., {Dalal}, N., {Holder}, G., {Kuhlen}, M., {Marrone}, D.,
  {Murray}, N., \& {Vieira}, J. 2013, \apj, 767, 9

\bibitem[{{Hurvich} \& {Tsai}(1989)}]{Hurvich++89}
{Hurvich}, C.~M., \& {Tsai}, C.-L. 1989, Biometrika, 76, 297

\bibitem[{{Kaufmann} {et~al.}(2008){Kaufmann}, {Bullock}, {Maller}, \&
  {Fang}}]{Kaufmann++08}
{Kaufmann}, T., {Bullock}, J.~S., {Maller}, A., \& {Fang}, T. 2008, in American
  Institute of Physics Conference Series, Vol. 1035, The Evolution of Galaxies
  Through the Neutral Hydrogen Window, ed. {R.~Minchin \& E.~Momjian}, 147--150

\bibitem[{{Kawano} {et~al.}(2004){Kawano}, {Oguri}, {Matsubara}, \&
  {Ikeuchi}}]{Kawano++04}
{Kawano}, Y., {Oguri}, M., {Matsubara}, T., \& {Ikeuchi}, S. 2004, \pasj, 56,
  253

\bibitem[{{Keeton}(2001{\natexlab{a}})}]{Keeton++01b}
{Keeton}, C.~R. 2001{\natexlab{a}}, ArXiv Astrophysics e-prints

\bibitem[{{Keeton}(2001{\natexlab{b}})}]{Keeton++01a}
---. 2001{\natexlab{b}}, ArXiv Astrophysics e-prints

\bibitem[{{Keeton}(2001{\natexlab{c}})}]{Keeton++01}
---. 2001{\natexlab{c}}, ArXiv Astrophysics e-prints

\bibitem[{{Keeton} {et~al.}(2006){Keeton}, {Burles}, {Schechter}, \&
  {Wambsganss}}]{Keeton++06}
{Keeton}, C.~R., {Burles}, S., {Schechter}, P.~L., \& {Wambsganss}, J. 2006,
  \apj, 639, 1

\bibitem[{{Keeton} \& {Moustakas}(2009)}]{K+M09}
{Keeton}, C.~R., \& {Moustakas}, L.~A. 2009, \apj, 699, 1720

\bibitem[{{Kelly} {et~al.}(2014){Kelly}, {Becker}, {Sobolewska},
  {Siemiginowska}, \& {Uttley}}]{Kelly++14}
{Kelly}, B.~C., {Becker}, A.~C., {Sobolewska}, M., {Siemiginowska}, A., \&
  {Uttley}, P. 2014, ArXiv e-prints

\bibitem[{{Klypin} {et~al.}(1999){Klypin}, {Kravtsov}, {Valenzuela}, \&
  {Prada}}]{Klypin++99}
{Klypin}, A., {Kravtsov}, A.~V., {Valenzuela}, O., \& {Prada}, F. 1999, \apj,
  522, 82

\bibitem[{{Kochanek} \& {Dalal}(2004)}]{Kochanek++04}
{Kochanek}, C.~S., \& {Dalal}, N. 2004, \apj, 610, 69

\bibitem[{{Komatsu} {et~al.}(2011){Komatsu}, {Smith}, {Dunkley}, {Bennett},
  {Gold}, {Hinshaw}, {Jarosik}, {Larson}, {Nolta}, {Page}, {Spergel},
  {Halpern}, {Hill}, {Kogut}, {Limon}, {Meyer}, {Odegard}, {Tucker}, {Weiland},
  {Wollack}, \& {Wright}}]{Komatsu++11}
{Komatsu}, E., {Smith}, K.~M., {Dunkley}, J., {Bennett}, C.~L., {Gold}, B.,
  {Hinshaw}, G., {Jarosik}, N., {Larson}, D., {Nolta}, M.~R., {Page}, L.,
  {Spergel}, D.~N., {Halpern}, M., {Hill}, R.~S., {Kogut}, A., {Limon}, M.,
  {Meyer}, S.~S., {Odegard}, N., {Tucker}, G.~S., {Weiland}, J.~L., {Wollack},
  E., \& {Wright}, E.~L. 2011, \apjs, 192, 18

\bibitem[{{Koopmans}(2005)}]{Koo05}
{Koopmans}, L.~V.~E. 2005, \mnras, 363, 1136

\bibitem[{{Koopmans} {et~al.}(2009{\natexlab{a}}){Koopmans}, {Bolton}, {Treu},
  {Czoske}, {Auger}, {Barnab{\`e}}, {Vegetti}, {Gavazzi}, {Moustakas}, \&
  {Burles}}]{Koo++09}
{Koopmans}, L.~V.~E., {Bolton}, A., {Treu}, T., {Czoske}, O., {Auger}, M.~W.,
  {Barnab{\`e}}, M., {Vegetti}, S., {Gavazzi}, R., {Moustakas}, L.~A., \&
  {Burles}, S. 2009{\natexlab{a}}, \apjl, 703, L51

\bibitem[{{Koopmans} {et~al.}(2009{\natexlab{b}}){Koopmans}, {Bolton}, {Treu},
  {Czoske}, {Auger}, {Barnab{\`e}}, {Vegetti}, {Gavazzi}, {Moustakas}, \&
  {Burles}}]{Koopmans++09}
---. 2009{\natexlab{b}}, \apjl, 703, L51

\bibitem[{{Koopmans} {et~al.}(2006){Koopmans}, {Treu}, {Bolton}, {Burles}, \&
  {Moustakas}}]{Koo++06}
{Koopmans}, L.~V.~E., {Treu}, T., {Bolton}, A.~S., {Burles}, S., \&
  {Moustakas}, L.~A. 2006, \apj, 649, 599

\bibitem[{{Koposov} {et~al.}(2008){Koposov}, {Belokurov}, {Evans}, {Hewett},
  {Irwin}, {Gilmore}, {Zucker}, {Rix}, {Fellhauer}, {Bell}, \&
  {Glushkova}}]{Koposov++08}
{Koposov}, S., {Belokurov}, V., {Evans}, N.~W., {Hewett}, P.~C., {Irwin},
  M.~J., {Gilmore}, G., {Zucker}, D.~B., {Rix}, H.-W., {Fellhauer}, M., {Bell},
  E.~F., \& {Glushkova}, E.~V. 2008, \apj, 686, 279

\bibitem[{{Kravtsov} {et~al.}(2004){Kravtsov}, {Berlind}, {Wechsler}, {Klypin},
  {Gottl{\"o}ber}, {Allgood}, \& {Primack}}]{Kravtsov++04}
{Kravtsov}, A.~V., {Berlind}, A.~A., {Wechsler}, R.~H., {Klypin}, A.~A.,
  {Gottl{\"o}ber}, S., {Allgood}, B., \& {Primack}, J.~R. 2004, \apj, 609, 35

\bibitem[{{Kundic} {et~al.}(1997){Kundic}, {Hogg}, {Blandford}, {Cohen},
  {Lubin}, \& {Larkin}}]{Kundic++97}
{Kundic}, T., {Hogg}, D.~W., {Blandford}, R.~D., {Cohen}, J.~G., {Lubin},
  L.~M., \& {Larkin}, J.~E. 1997, \aj, 114, 2276

\bibitem[{{Lagattuta} {et~al.}(2010){Lagattuta}, {Fassnacht}, {Auger},
  {Marshall}, {Brada{\v c}}, {Treu}, {Gavazzi}, {Schrabback}, {Faure}, \&
  {Anguita}}]{Lagattuta++10}
{Lagattuta}, D.~J., {Fassnacht}, C.~D., {Auger}, M.~W., {Marshall}, P.~J.,
  {Brada{\v c}}, M., {Treu}, T., {Gavazzi}, R., {Schrabback}, T., {Faure}, C.,
  \& {Anguita}, T. 2010, \apj, 716, 1579

\bibitem[{{Larkin} {et~al.}(2006){Larkin}, {Barczys}, {Krabbe}, {Adkins},
  {Aliado}, {Amico}, {Brims}, {Campbell}, {Canfield}, {Gasaway}, {Honey},
  {Iserlohe}, {Johnson}, {Kress}, {LaFreniere}, {Magnone}, {Magnone},
  {McElwain}, {Moon}, {Quirrenbach}, {Skulason}, {Song}, {Spencer}, {Weiss}, \&
  {Wright}}]{Larkin++06}
{Larkin}, J., {Barczys}, M., {Krabbe}, A., {Adkins}, S., {Aliado}, T., {Amico},
  P., {Brims}, G., {Campbell}, R., {Canfield}, J., {Gasaway}, T., {Honey}, A.,
  {Iserlohe}, C., {Johnson}, C., {Kress}, E., {LaFreniere}, D., {Magnone}, K.,
  {Magnone}, N., {McElwain}, M., {Moon}, J., {Quirrenbach}, A., {Skulason}, G.,
  {Song}, I., {Spencer}, M., {Weiss}, J., \& {Wright}, S. 2006, \nar, 50, 362

\bibitem[{{Li} {et~al.}(2014){Li}, {Shan}, {Mo}, {Kneib}, {Yang}, {Luo}, {van
  den Bosch}, {Erben}, {Moraes}, \& {Makler}}]{Li++14}
{Li}, R., {Shan}, H., {Mo}, H., {Kneib}, J.-P., {Yang}, X., {Luo}, W., {van den
  Bosch}, F.~C., {Erben}, T., {Moraes}, B., \& {Makler}, M. 2014, \mnras, 438,
  2864

\bibitem[{{Lu} {et~al.}(2012){Lu}, {Mo}, {Katz}, \& {Weinberg}}]{Lu++12}
{Lu}, Y., {Mo}, H.~J., {Katz}, N., \& {Weinberg}, M.~D. 2012, \mnras, 421, 1779

\bibitem[{{Macci{\`o}}(2010)}]{Maccio++10}
{Macci{\`o}}, A.~V. 2010, in American Institute of Physics Conference Series,
  Vol. 1240, American Institute of Physics Conference Series, ed.
  {V.~P.~Debattista \& C.~C.~Popescu}, 355--358

\bibitem[{{Macci{\`o}} {et~al.}(2008){Macci{\`o}}, {Dutton}, \& {van den
  Bosch}}]{Maccio++08}
{Macci{\`o}}, A.~V., {Dutton}, A.~A., \& {van den Bosch}, F.~C. 2008, \mnras,
  391, 1940

\bibitem[{{MacLeod} {et~al.}(2013){MacLeod}, {Jones}, {Agol}, \&
  {Kochanek}}]{MacLeod++13}
{MacLeod}, C.~L., {Jones}, R., {Agol}, E., \& {Kochanek}, C.~S. 2013, \apj,
  773, 35

\bibitem[{{MacLeod} {et~al.}(2009){MacLeod}, {Kochanek}, \&
  {Agol}}]{MacLeod++09}
{MacLeod}, C.~L., {Kochanek}, C.~S., \& {Agol}, E. 2009, \apj, 699, 1578

\bibitem[{{Mao} \& {Schneider}(1998)}]{Mao++98}
{Mao}, S., \& {Schneider}, P. 1998, \mnras, 295, 587

\bibitem[{{Max} {et~al.}(2008){Max}, {McGrath}, {Gavel}, {Le Mignant},
  {Wizinowich}, \& {Dekany}}]{Max++08}
{Max}, C., {McGrath}, E., {Gavel}, D., {Le Mignant}, D., {Wizinowich}, P., \&
  {Dekany}, R. 2008, in Society of Photo-Optical Instrumentation Engineers
  (SPIE) Conference Series, Vol. 7015, Society of Photo-Optical Instrumentation
  Engineers (SPIE) Conference Series

\bibitem[{{McCully} {et~al.}(2014){McCully}, {Keeton}, {Wong}, \&
  {Zabludoff}}]{McCully++2014}
{McCully}, C., {Keeton}, C.~R., {Wong}, K.~C., \& {Zabludoff}, A.~I. 2014,
  ArXiv e-prints

\bibitem[{{McKean} {et~al.}(2007){McKean}, {Koopmans}, {Flack}, {Fassnacht},
  {Thompson}, {Matthews}, {Blandford}, {Readhead}, \& {Soifer}}]{McKean++07}
{McKean}, J.~P., {Koopmans}, L.~V.~E., {Flack}, C.~E., {Fassnacht}, C.~D.,
  {Thompson}, D., {Matthews}, K., {Blandford}, R.~D., {Readhead}, A.~C.~S., \&
  {Soifer}, B.~T. 2007, \mnras, 378, 109

\bibitem[{{Menci} {et~al.}(2012){Menci}, {Fiore}, \& {Lamastra}}]{Menci++12}
{Menci}, N., {Fiore}, F., \& {Lamastra}, A. 2012, \mnras, 421, 2384

\bibitem[{{Metcalf} \& {Amara}(2012)}]{Metcalf++12}
{Metcalf}, R.~B., \& {Amara}, A. 2012, \mnras, 419, 3414

\bibitem[{{Metcalf} \& {Madau}(2001{\natexlab{a}})}]{M+M01}
{Metcalf}, R.~B., \& {Madau}, P. 2001{\natexlab{a}}, \apj, 563, 9

\bibitem[{{Metcalf} \& {Madau}(2001{\natexlab{b}})}]{Metcalf++01}
---. 2001{\natexlab{b}}, \apj, 563, 9

\bibitem[{{Metcalf} {et~al.}(2004){Metcalf}, {Moustakas}, {Bunker}, \&
  {Parry}}]{Metcalf++04}
{Metcalf}, R.~B., {Moustakas}, L.~A., {Bunker}, A.~J., \& {Parry}, I.~R. 2004,
  \apj, 607, 43

\bibitem[{{Moore} {et~al.}(1999){Moore}, {Ghigna}, {Governato}, {Lake},
  {Quinn}, {Stadel}, \& {Tozzi}}]{Moore++1999}
{Moore}, B., {Ghigna}, S., {Governato}, F., {Lake}, G., {Quinn}, T., {Stadel},
  J., \& {Tozzi}, P. 1999, \apjl, 524, L19

\bibitem[{{Moustakas} \& {Metcalf}(2003)}]{Moustakas++03}
{Moustakas}, L.~A., \& {Metcalf}, R.~B. 2003, \mnras, 339, 607

\bibitem[{{Mu{\~n}oz} {et~al.}(2001){Mu{\~n}oz}, {Kochanek}, \&
  {Keeton}}]{Munoz++01}
{Mu{\~n}oz}, J.~A., {Kochanek}, C.~S., \& {Keeton}, C.~R. 2001, \apj, 558, 657

\bibitem[{{M{\"u}ller-S{\'a}nchez} {et~al.}(2011){M{\"u}ller-S{\'a}nchez},
  {Prieto}, {Hicks}, {Vives-Arias}, {Davies}, {Malkan}, {Tacconi}, \&
  {Genzel}}]{Muller++11}
{M{\"u}ller-S{\'a}nchez}, F., {Prieto}, M.~A., {Hicks}, E.~K.~S.,
  {Vives-Arias}, H., {Davies}, R.~I., {Malkan}, M., {Tacconi}, L.~J., \&
  {Genzel}, R. 2011, \apj, 739, 69

\bibitem[{{Murayama} {et~al.}(1999){Murayama}, {Taniguchi}, {Evans}, {Sanders},
  {Hodapp}, {Kawara}, \& {Arimoto}}]{Murayama++1999}
{Murayama}, T., {Taniguchi}, Y., {Evans}, A.~S., {Sanders}, D.~B., {Hodapp},
  K.-W., {Kawara}, K., \& {Arimoto}, N. 1999, \aj, 117, 1645

\bibitem[{{Navarro} {et~al.}(1996){Navarro}, {Frenk}, \& {White}}]{NFW++1996}
{Navarro}, J.~F., {Frenk}, C.~S., \& {White}, S.~D.~M. 1996, \apj, 462, 563

\bibitem[{{Nierenberg} {et~al.}(2013){Nierenberg}, {Treu}, {Menci}, {Lu}, \&
  {Wang}}]{Nierenberg++13}
{Nierenberg}, A.~M., {Treu}, T., {Menci}, N., {Lu}, Y., \& {Wang}, W. 2013,
  ArXiv e-prints

\bibitem[{Oguri \& Marshall(2010)}]{Oguri++10}
Oguri, M., \& Marshall, P.~J. 2010, \mnras, 405, 2579

\bibitem[{{Oke}(1974)}]{Oke++1974}
{Oke}, J.~B. 1974, \apjs, 27, 21

\bibitem[{{Papastergis} {et~al.}(2011){Papastergis}, {Martin}, {Giovanelli}, \&
  {Haynes}}]{Papastergis++11}
{Papastergis}, E., {Martin}, A.~M., {Giovanelli}, R., \& {Haynes}, M.~P. 2011,
  \apj, 739, 38

\bibitem[{{Patnaik} {et~al.}(1992){Patnaik}, {Browne}, {Walsh}, {Chaffee}, \&
  {Foltz}}]{Patnaik++92}
{Patnaik}, A.~R., {Browne}, I.~W.~A., {Walsh}, D., {Chaffee}, F.~H., \&
  {Foltz}, C.~B. 1992, \mnras, 259, 1P

\bibitem[{{Patnaik} {et~al.}(1999){Patnaik}, {Kemball}, {Porcas}, \&
  {Garrett}}]{Patnaik++1999}
{Patnaik}, A.~R., {Kemball}, A.~J., {Porcas}, R.~W., \& {Garrett}, M.~A. 1999,
  \mnras, 307, L1

\bibitem[{{Planck Collaboration} {et~al.}(2013){Planck Collaboration}, {Ade},
  {Aghanim}, {Armitage-Caplan}, {Arnaud}, {Ashdown}, {Atrio-Barandela},
  {Aumont}, {Baccigalupi}, {Banday}, \& et~al.}]{Planck++13}
{Planck Collaboration}, {Ade}, P.~A.~R., {Aghanim}, N., {Armitage-Caplan}, C.,
  {Arnaud}, M., {Ashdown}, M., {Atrio-Barandela}, F., {Aumont}, J.,
  {Baccigalupi}, C., {Banday}, A.~J., \& et~al. 2013, ArXiv e-prints

\bibitem[{{Rocha} {et~al.}(2013){Rocha}, {Peter}, {Bullock}, {Kaplinghat},
  {Garrison-Kimmel}, {O{\~n}orbe}, \& {Moustakas}}]{Rocha++13}
{Rocha}, M., {Peter}, A.~H.~G., {Bullock}, J.~S., {Kaplinghat}, M.,
  {Garrison-Kimmel}, S., {O{\~n}orbe}, J., \& {Moustakas}, L.~A. 2013, \mnras,
  430, 81

\bibitem[{{Sluse} {et~al.}(2012{\natexlab{a}}){Sluse}, {Chantry}, {Magain},
  {Courbin}, \& {Meylan}}]{Sluse++12b}
{Sluse}, D., {Chantry}, V., {Magain}, P., {Courbin}, F., \& {Meylan}, G.
  2012{\natexlab{a}}, \aap, 538, A99

\bibitem[{{Sluse} {et~al.}(2012{\natexlab{b}}){Sluse}, {Hutsem{\'e}kers},
  {Courbin}, {Meylan}, \& {Wambsganss}}]{Sluse++12a}
{Sluse}, D., {Hutsem{\'e}kers}, D., {Courbin}, F., {Meylan}, G., \&
  {Wambsganss}, J. 2012{\natexlab{b}}, \aap, 544, A62

\bibitem[{{Somerville}(2002)}]{Somerville++02}
{Somerville}, R.~S. 2002, \apjl, 572, L23

\bibitem[{{Springel}(2010)}]{Springel++10}
{Springel}, V. 2010, \araa, 48, 391

\bibitem[{{Strigari} {et~al.}(2007){Strigari}, {Bullock}, {Kaplinghat},
  {Diemand}, {Kuhlen}, \& {Madau}}]{Strigari++07}
{Strigari}, L.~E., {Bullock}, J.~S., {Kaplinghat}, M., {Diemand}, J., {Kuhlen},
  M., \& {Madau}, P. 2007, \apj, 669, 676

\bibitem[{{Sugai} {et~al.}(2007){Sugai}, {Kawai}, {Shimono}, {Hattori},
  {Kosugi}, {Kashikawa}, {Inoue}, \& {Chiba}}]{Sugai++07}
{Sugai}, H., {Kawai}, A., {Shimono}, A., {Hattori}, T., {Kosugi}, G.,
  {Kashikawa}, N., {Inoue}, K.~T., \& {Chiba}, M. 2007, \apj, 660, 1016

\bibitem[{{Tollerud} {et~al.}(2008){Tollerud}, {Bullock}, {Strigari}, \&
  {Willman}}]{Toll++08}
{Tollerud}, E.~J., {Bullock}, J.~S., {Strigari}, L.~E., \& {Willman}, B. 2008,
  \apj, 688, 277

\bibitem[{{Treu}(2010)}]{Treu++10}
{Treu}, T. 2010, \araa, 48, 87

\bibitem[{{van Dam} {et~al.}(2006){van Dam}, {Bouchez}, {Le Mignant},
  {Johansson}, {Wizinowich}, {Campbell}, {Chin}, {Hartman}, {Lafon}, {Stomski},
  \& {Summers}}]{vanDam++06}
{van Dam}, M.~A., {Bouchez}, A.~H., {Le Mignant}, D., {Johansson}, E.~M.,
  {Wizinowich}, P.~L., {Campbell}, R.~D., {Chin}, J.~C.~Y., {Hartman}, S.~K.,
  {Lafon}, R.~E., {Stomski}, Jr., P.~J., \& {Summers}, D.~M. 2006, \pasp, 118,
  310

\bibitem[{{Vegetti} {et~al.}(2010{\natexlab{a}}){Vegetti}, {Czoske}, \&
  {Koopmans}}]{Vegetti++10b}
{Vegetti}, S., {Czoske}, O., \& {Koopmans}, L.~V.~E. 2010{\natexlab{a}},
  \mnras, 407, 225

\bibitem[{{Vegetti} {et~al.}(2010{\natexlab{b}}){Vegetti}, {Koopmans},
  {Bolton}, {Treu}, \& {Gavazzi}}]{Vegetti++10a}
{Vegetti}, S., {Koopmans}, L.~V.~E., {Bolton}, A., {Treu}, T., \& {Gavazzi}, R.
  2010{\natexlab{b}}, \mnras, 408, 1969

\bibitem[{{Vegetti} {et~al.}(2012){Vegetti}, {Lagattuta}, {McKean}, {Auger},
  {Fassnacht}, \& {Koopmans}}]{Vegetti++12}
{Vegetti}, S., {Lagattuta}, D.~J., {McKean}, J.~P., {Auger}, M.~W.,
  {Fassnacht}, C.~D., \& {Koopmans}, L.~V.~E. 2012, \nat, 481, 341

\bibitem[{{Vegetti et al.}(2014)}]{Vegetti++14}
{Vegetti et al.}, S. 2014, MNRAS, submitted

\bibitem[{{Walker} \& {Pe{\~n}arrubia}(2011)}]{Walker++11}
{Walker}, M.~G., \& {Pe{\~n}arrubia}, J. 2011, \apj, 742, 20

\bibitem[{{Wizinowich} {et~al.}(2006){Wizinowich}, {Le Mignant}, {Bouchez},
  {Campbell}, {Chin}, {Contos}, {van Dam}, {Hartman}, {Johansson}, {Lafon},
  {Lewis}, {Stomski}, {Summers}, {Brown}, {Danforth}, {Max}, \&
  {Pennington}}]{Wizinowich++06}
{Wizinowich}, P.~L., {Le Mignant}, D., {Bouchez}, A.~H., {Campbell}, R.~D.,
  {Chin}, J.~C.~Y., {Contos}, A.~R., {van Dam}, M.~A., {Hartman}, S.~K.,
  {Johansson}, E.~M., {Lafon}, R.~E., {Lewis}, H., {Stomski}, P.~J., {Summers},
  D.~M., {Brown}, C.~G., {Danforth}, P.~M., {Max}, C.~E., \& {Pennington},
  D.~M. 2006, \pasp, 118, 297

\bibitem[{{Wolf} \& {Bullock}(2012)}]{Wolf++12}
{Wolf}, J., \& {Bullock}, J.~S. 2012, ArXiv e-prints

\bibitem[{{Xu} {et~al.}(2012){Xu}, {Mao}, {Cooper}, {Gao}, {Frenk}, {Angulo},
  \& {Helly}}]{Xu++12}
{Xu}, D.~D., {Mao}, S., {Cooper}, A.~P., {Gao}, L., {Frenk}, C.~S., {Angulo},
  R.~E., \& {Helly}, J. 2012, \mnras, 421, 2553

\bibitem[{{Xu} {et~al.}(2013){Xu}, {Sluse}, {Gao}, {Wang}, {Frenk}, {Mao}, \&
  {Schneider}}]{Xu++14}
{Xu}, D.~D., {Sluse}, D., {Gao}, L., {Wang}, J., {Frenk}, C., {Mao}, S., \&
  {Schneider}, P. 2013, ArXiv e-prints

\bibitem[{{Yoo} {et~al.}(2005){Yoo}, {Kochanek}, {Falco}, \&
  {McLeod}}]{Yoo++05}
{Yoo}, J., {Kochanek}, C.~S., {Falco}, E.~E., \& {McLeod}, B.~A. 2005, \apj,
  626, 51

\bibitem[{{Yoo} {et~al.}(2006){Yoo}, {Kochanek}, {Falco}, \&
  {McLeod}}]{Yoo++06}
---. 2006, \apj, 642, 22

\bibitem[{{Zolotov} {et~al.}(2012){Zolotov}, {Brooks}, {Willman}, {Governato},
  {Pontzen}, {Christensen}, {Dekel}, {Quinn}, {Shen}, \&
  {Wadsley}}]{Zolotov++12}
{Zolotov}, A., {Brooks}, A.~M., {Willman}, B., {Governato}, F., {Pontzen}, A.,
  {Christensen}, C., {Dekel}, A., {Quinn}, T., {Shen}, S., \& {Wadsley}, J.
  2012, \apj, 761, 71

\end{thebibliography}
\appendix
\section{Variation in lensing effect with perturber mass profile}

\begin{figure*}
\centering
\includegraphics[scale=0.35,  trim = 0 0 40 20, clip = true]{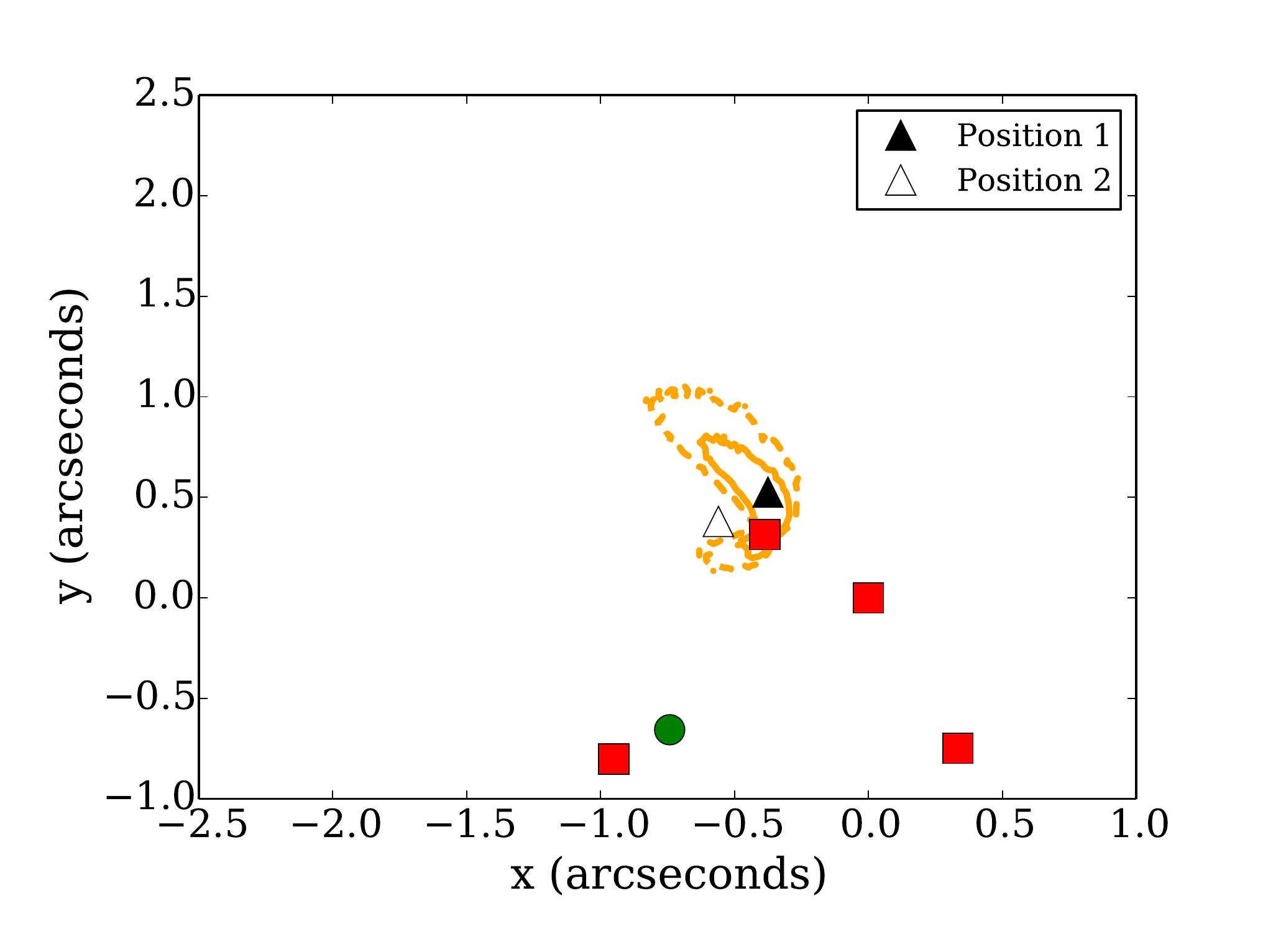} \includegraphics[scale=0.35,  trim = 0 0 30 20, clip = true]{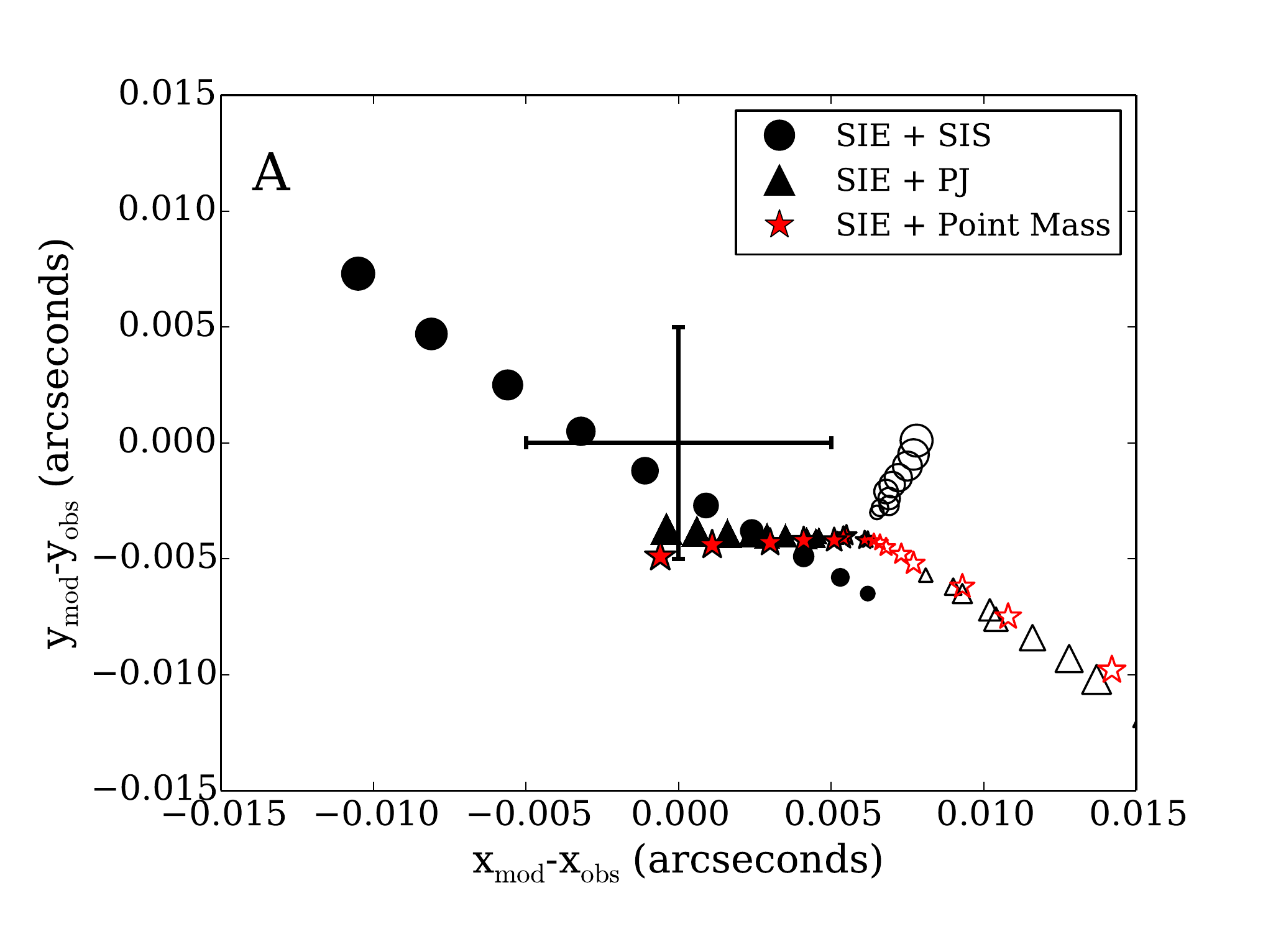} \includegraphics[scale=0.35,  trim = 0 0 30 20, clip = true]{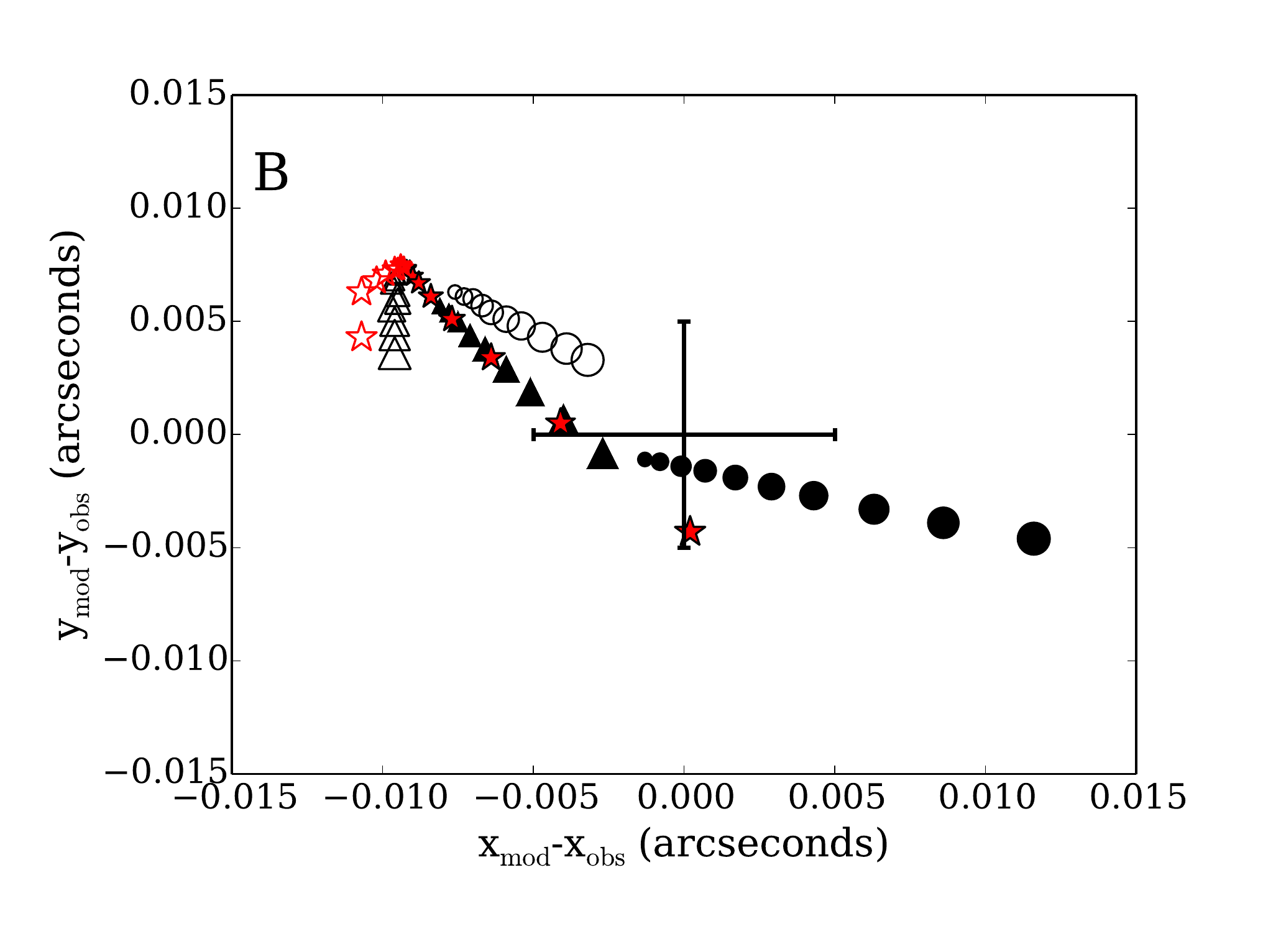}  \includegraphics[scale=0.35,  trim = 0 0 30 20, clip = true]{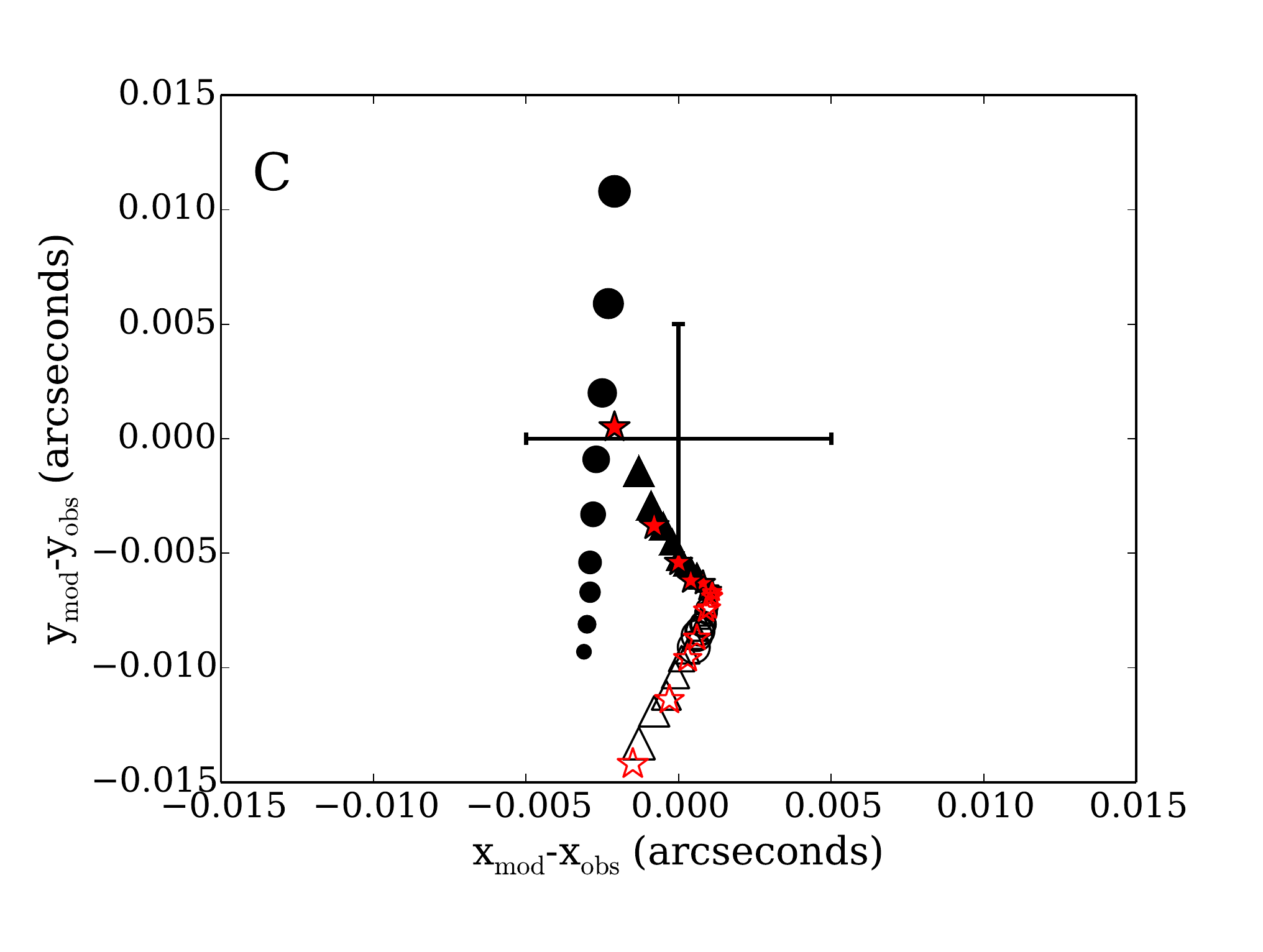}

\caption{Demonstration of how the model image positions vary relative to the observed image positions as the perturber Einstein radius is varied logarithmically for fixed perturber position and SIE macromodel parameters, for an SIS, PJ and point mass perturber represented by circles, triangles, and red stars respectively. Point sizes increase with increasing perturber Einstein radius. We explore two perturber positions, illustrated in the upper left panel by open and filled triangles, where the red squares and green circle represent observed image and galaxy positions respectively and the solid and dashed contours represent the 68\% and 95\% confidence interval for the position of the perturber in the case of the PJ mass profile.}
\label{fig:testPos}
\end{figure*}

In this appendix we illustrate the difference in the posterior probability distributions for the perturber position and mass seen in Figure \ref{fig:prmass} in the case of the PJ and SIS perturber, by means of a simple example. The most notable difference between the two models is the large gap to the East of image A in the case of the PJ perturber. In order to clarify the origin of this difference, in Figure \ref{fig:testPos} we show the model prediction for the image positions as the perturber Einstein radius is varied for fixed position and host macromodel, in the case that the perturber is in the gap (position 2), and also in the region further North which is permitted for both mass profiles (position 1). We also show the results in the case of a point mass perturber, which has no convergence at all at the image positions. This is the extreme opposite to the SIS, which has a density profile which extends outward indefinitely.
 The perturber positions relative to the lensed images are given in the lower left panel of Figure \ref{fig:testPos}. 


The first thing to note in this figure is that the perturber in position 1 does not provide a good fit to the observation in either the case of the SIS or PJ perturber (the best fit $\chi^2$ per degree of freedom is approximately 8 in the case of the SIS perturber, and $\sim 50$ in the case of the PJ perturber). This figure illustrates the fundamental difference in the lensing effect between the truncated and non-truncated perturber profiles, due to the fact that the lensing signal in the case of the truncated profile is dominated by shear, while the SIS profile has significant convergence at the image position. In fact, the PJ perturber has an almost identical effect to the point mass perturber. In position 2, the PJ profile simply cannot provide a good fit to the astrometry, irrespective of its mass.

\begin{figure*}
\centering
\includegraphics[scale=0.35]{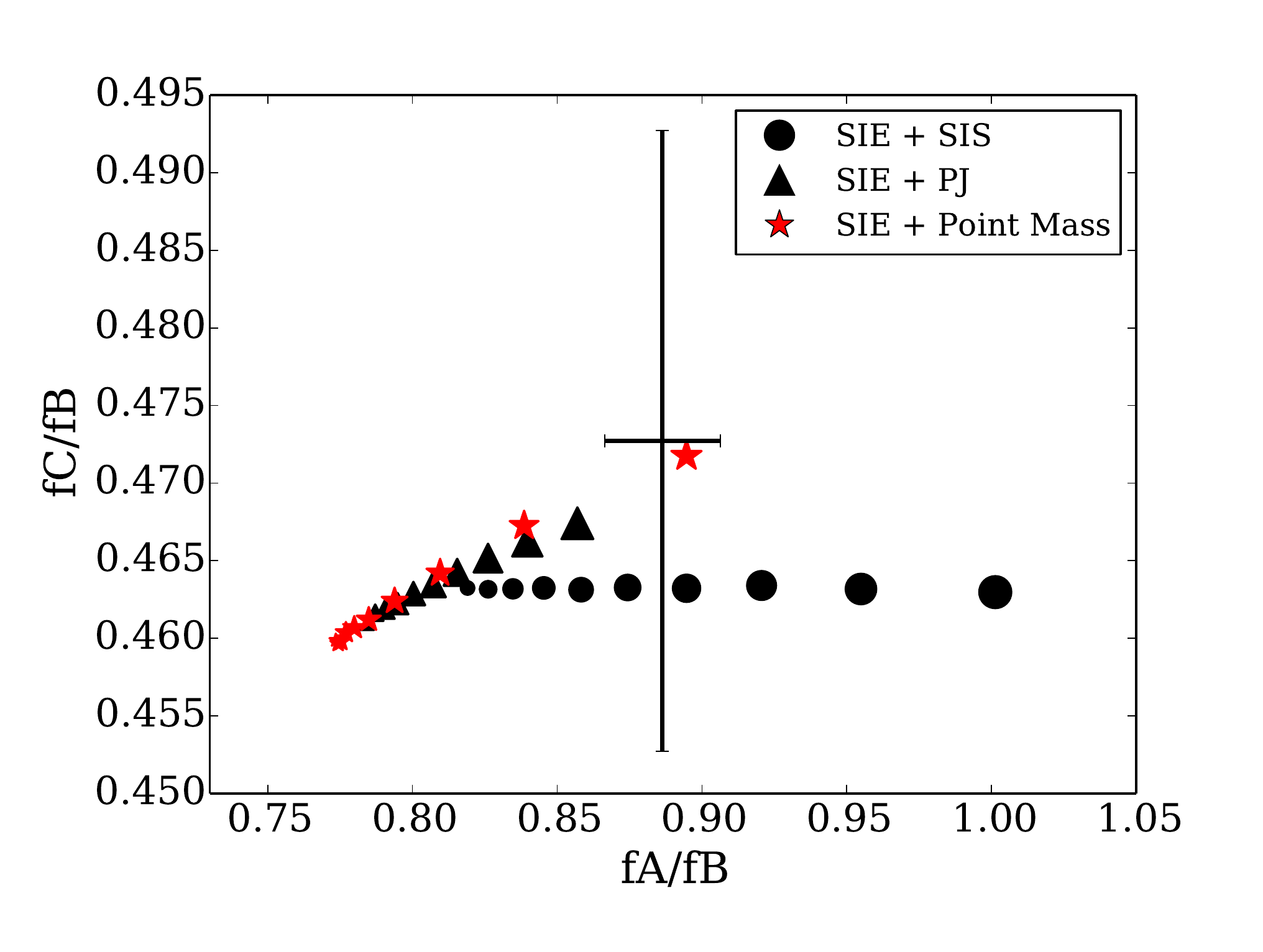}\includegraphics[scale=0.35]{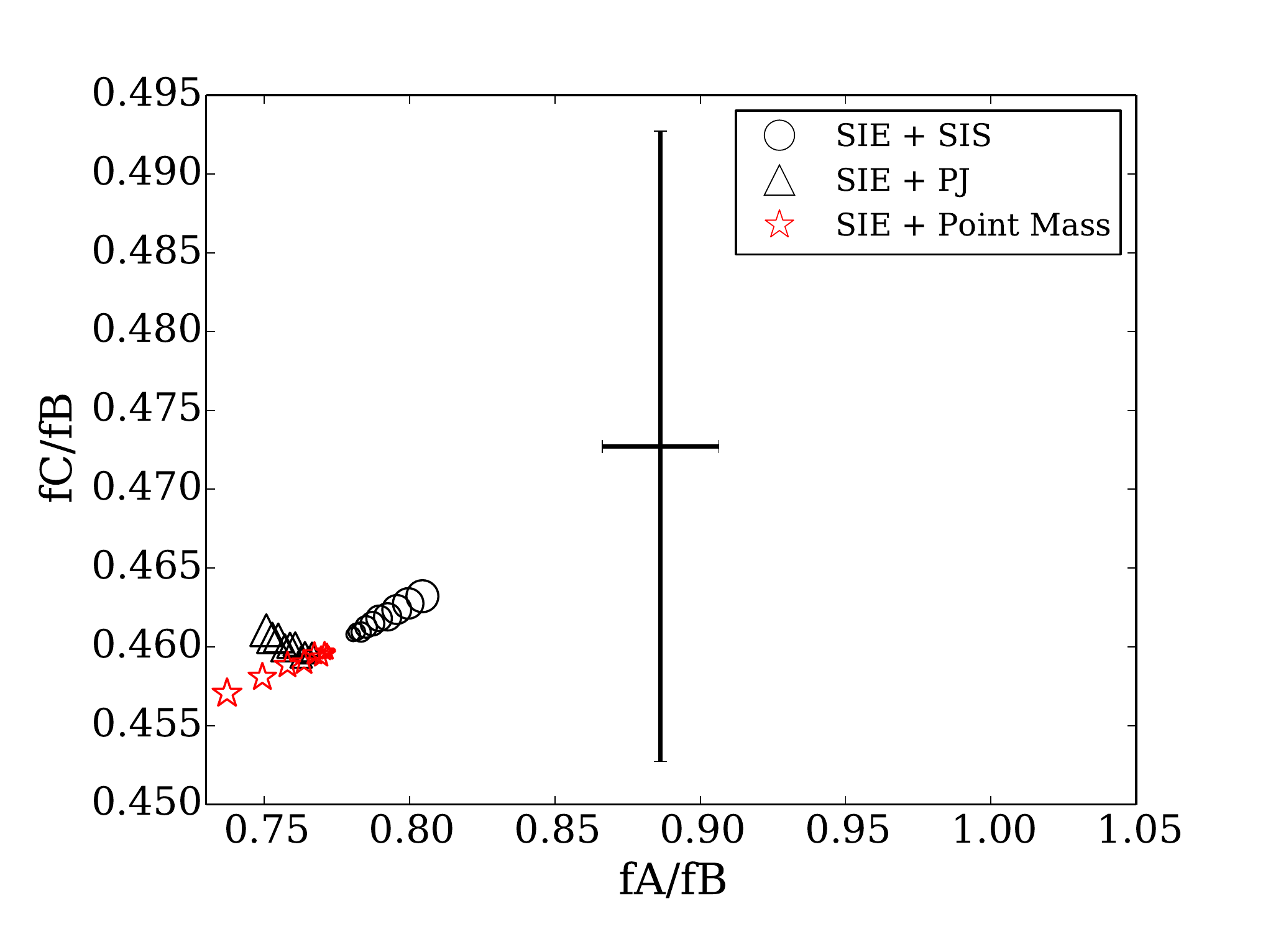}
\caption{Model flux ratios as the perturber mass is varied for fixed position for two different mass profiles in the same way as in Figure \ref{fig:testPos}. For clarity, we show results for positions one and two separately.}
\label{fig:testFlux}
\end{figure*}

In Figure \ref{fig:testFlux} we plot the resulting flux ratios as the perturber masses are varied in the three cases. As before, we see that in the three cases, the perturber in position 2 provides a less good fit to the observation in the case of the SIS. The fit is much worse though in the case of the PJ and point mass perturbers which again show almost identical behaviour. Unlike the case of the SIS profile, in position 2 the flux of image A relative to image B does not increase significantly as the perturber mass is increased for the PJ perturber. As expected, the flux ratio between image B and C does not vary significantly as the perturber mass is varied, since the dominant effect is to alter the flux of image A.


\label{lastpage}
\bsp
\end{document}